\documentclass{JINST}

\usepackage{graphicx}

\usepackage{amssymb}

\usepackage{threeparttable}

\newcommand{\antinu}{anti-neutrino}
\newcommand{\ls}{LS}
\newcommand{\sixplusonew}{6+1W}
\newcommand{\fiveplusonew}{5+1W}
\newcommand{\fiveplusone}{5+1}

\newcommand{\fourplusone}{4+1}
\newcommand{\threeplusone}{3+1}
\newcommand{\composite}{$^{68}$Ge$^{60}$Co}
\newcommand{\lead}{$^{210}$Pb}
\newcommand{\radon}{$^{222}$Rn}

\newcommand{\ius}{IUs}
\newcommand{\iu}{IU}

\def\@fnsymbol#1{\ifcase#1\hbox{}\or *\or \dagger\or \ddagger\or \mathchar "278\or \mathchar "27B\or \|\or **\or \dagger\dagger \or \ddagger\ddagger \or \mathchar"27C \else\@ctrerr\fi\relax}

\title{The KamLAND Full-Volume Calibration System}

\author{B.E.~Berger$^a$, 
J.~Busenitz$^b$,
T.~Classen$^{b}$\thanks{Present address: Department of Physics, University of California, Davis, CA 95616, USA},
M.P.~Decowski$^{c,g}$,
D.A.~Dwyer$^d$,
G.~Elor$^c$,
A.~Frank$^c$,
S.J.~Freedman$^{c,g}$,
B.K.~Fujikawa$^{c,g}$,
M.~Galloway$^c$,
F.~Gray$^{c}$\thanks{Present address:  Department of Physics and Computational Science, Regis University, Denver, CO 80221, USA},
K.M.~Heeger$^{e,g}$,
L.~Hsu$^{c}$\thanks{Present address: Fermi National Accelerator Laboratory, Batavia, IL 60510, USA},
K.~Ichimura$^c$, 
R.~Kadel$^c$,
G.~Keefer$^b$,
C.~Lendvai$^c$,
D.~McKee$^{b}$\thanks{Present address: Department of Physics, Kansas State University,  Manhattan, KS 66506, USA},
T.~O'Donnell$^c$,
A.~Piepke$^{b,g}$,
H.M.~Steiner$^c$,
D.~Syversrud$^c$,
J.~Wallig$^c$,
L.A.~Winslow$^{c}$\thanks{Present address: Department of Physics, Massachusetts Institute of Technology, Cambridge, MA 02139, USA},
T.~Ebihara$^f$, 
S.~Enomoto$^{f,g}$, 
K.~Furuno$^f$, 
Y.~Gando$^f$, 
H.~Ikeda$^f$, 
K.~Inoue$^{f,g}$, 
Y.~Kibe$^f$, 
Y.~Kishimoto$^f$,
M.~Koga$^{f,g}$,
Y.~Minekawa$^f$,
T.~Mitsui$^f$,
K.~Nakajima$^{f}$\thanks{Present address: Center of Quantum Universe, Okayama University, Okayama 700-8530, Japan},
K.~Nakajima$^f$,
K.~Nakamura$^{f,g}$,
K.~Owada$^f$,
I.~Shimizu$^f$,
Y.~Shimizu$^f$,
J.~Shirai$^f$,
F.~Suekane$^f$,
A.~Suzuki$^f$,
K.~Tamae$^f$,
S.~Yoshida$^f$,
A.~Kozlov$^g$,
H.~Murayama$^{g,c}$,
C.~Grant$^b$,
D.S.~Leonard$^{b}$\thanks{Present address: Department of Physics, University of Maryland, College Park, MD 20742, USA},
K.-B.~Luk$^c$,
C.~Jillings$^{d}$\thanks{Present address: SNOLAB, Lively, ON P3Y 1M3, Canada },
C.~Mauger$^{d}$\thanks{Present address: Subatomic Physics Group, Los Alamos National Laboratory, Los Alamos, NM 87545 \newline 
\indent{$^{\alpha}$Present address: Lawrence Livermore National Laboratory, Livermore, CA 94550, USA} \newline
\indent{$^{\beta}$ Present address: CENPA, University of Washington, Seattle, WA 98195, USA}
},
R.D.~McKeown$^d$,
C.~Zhang$^d$,
C.E.~Lane$^h$,
J.~Maricic$^h$,
T.~Miletic$^h$,
M.~Batygov$^i$,
J.G.~Learned$^i$,
S.~Matsuno$^i$,
S.~Pakvasa$^i$,
J.~Foster$^j$,
G.A.~Horton-Smith$^{j,g}$,
A.~Tang$^j$,
S.~Dazeley$^{k,\alpha}$,
K.E.~Downum$^m$,
G.~Gratta$^m$,
K.~Tolich$^{m,\beta}$,
W.~Bugg$^n$,
Y.~Efremenko$^{n,g}$,
Y.~Kamyshkov$^n$,
O.~Perevozchikov$^n$,
H.J.~Karwowski$^o$,
D.M.~Markoff$^o$,
W.~Tornow$^o$,
F.~Piquemal$^p$,
J.-S.~Ricol$^p$ \\
\llap{$^a$}{Department of Physics, Colorado  State University, \\
Fort Collins, Colorado 80523, USA} \\
\llap{$^b$}{Department of Physics and Astronomy, University of Alabama, \\
Tuscaloosa, Alabama 35487, USA} \\
\llap{$^c$}{Physics Department, University of  California, Berkeley and \\
Lawrence Berkeley National Laboratory, Berkeley, California 94720, USA} \\
\llap{$^d$}{W.~K.~Kellogg Radiation Laboratory, California Institute of Technology, \\
Pasadena, California 91125, USA} \\
\llap{$^e$}{Department of Physics, University of Wisconsin, \\
Madison, Wisconsin 53706, USA} \\
\llap{$^f$}{Research Center for Neutrino Science, Tohoku University, \\
Sendai 980-8578, Japan} \\
\llap{$^g$}{Institute for the Physics and Mathematics of the Universe, University of Tokyo, \\
Kashiwa, Japan 277-8568}     \\
\llap{$^h$}{Physics Department, Drexel University, \\
Philadelphia, Pennsylvania 19104, USA} \\
\llap{$^i$}{Department of Physics and Astronomy,
    University of Hawaii at Manoa, \\
    Honolulu, Hawaii 96822, USA} \\
\llap{$^j$}{Department of Physics,
    Kansas State University, \\
    Manhattan, Kansas 66506, USA} \\
\llap{$^k$}{Department of Physics and Astronomy,
    Louisiana State University, \\
    Baton Rouge, Louisiana 70803, USA} \\
\llap{$^m$}{Physics Department, Stanford
    University, \\
    Stanford, California 94305, USA} \\
\llap{$^n$}{Department of Physics and
    Astronomy, University of Tennessee, \\
    Knoxville, Tennessee 37996, USA} \\
\llap{$^o$}{Triangle Universities Nuclear
    Laboratory, Durham, North Carolina 27708, USA and \\
Physics Departments at Duke University, North Carolina Central University,
and the University of North Carolina at Chapel Hill} \\
\llap{$^p$}{CEN Bordeaux-Gradignan, IN2P3-CNRS and
    University Bordeaux I, \\
    F-33175 Gradignan Cedex, France} \\
}
\abstract{

We have successfully built and operated a source deployment system for the KamLAND detector.   This system was used to position radioactive sources throughout the delicate 1-kton liquid scintillator volume, while meeting stringent material cleanliness, material compatibility, and safety requirements.  The calibration data obtained with this device were used to fully characterize detector position and energy reconstruction biases.  As a result, the uncertainty in the size of the detector fiducial volume was reduced by a factor of two.  Prior to calibration with this system, the fiducial volume was the largest source of systematic uncertainty in measuring the number of \antinu s detected by KamLAND. This paper describes the design, operation and performance of this unique calibration system. }


\preprint{Date: \today}

\begin{document}


The Kamioka Liquid scintillator Anti-Neutrino Detector (KamLAND) is located in a mine operated by the Kamioka Mining and Smelting Company, Kamioka-cho, Gifu, Japan.  It was designed to study the flux of reactor \antinu s from nuclear power plants in Japan. In recent years, KamLAND made the first observation of the disappearance of   \antinu s from reactors.  It subsequently found evidence for \antinu\ oscillation in the distortion of the measured positron energy spectrum, and made a precision measurement of neutrino oscillation parameters \cite{KLprl1,KLprl2,KLprl3}. These observations have been critical in establishing that neutrinos oscillate and that the Large-Mixing-Angle Mikheyev-Smirnov-Wolfenstein (LMA-MSW) effect is the correct solution to the solar neutrino problem. 

KamLAND is a monolithic calorimeter consisting of 1~kton of ultra-pure liquid scintillator (LS).   Encased in a 13~m diameter, 135~$\mu$m thick, nylon/EVOH (ethylene vinyl alcohol copolymer) balloon, the LS volume serves as both the target and the detecting medium.  Kevlar ropes suspend the balloon in a bath of mineral oil, which is contained inside an 18~m diameter stainless-steel sphere. There are 1879 50-cm diameter photomultiplier tubes (PMTs) mounted on the inside of the stainless-steel sphere to detect the light from \antinu s interacting with the LS. 


Anti-neutrinos are detected via the inverse $\beta$-decay reaction, ${\bar\nu}_e + p \to e^+ + n$, in the hydrogen-rich scintillator target.  This interaction results in a time coincidence between two light-emitting events: a prompt $e^{+}$ which quickly annihilates, followed by a delayed 2.2~MeV gamma ray from the neutron capture on hydrogen. 

To avoid high levels of background near the balloon surface, the analysis region of the detector is defined using the reconstructed position of events in the final data analysis.  Cuts on the reconstructed position define a smaller target volume (the fiducial volume) that is used in the measurement of the \antinu\ flux and interaction rate per target. Systematic effects in the reconstruction of the event position lead to errors when defining the fiducial volume.  In previous analyses, resulting systematic uncertainty dominated the error in determining the expected rate of \antinu s\cite{KLprl1,KLprl2}. 

\begin{figure}[thpb]
\begin{minipage}[b]{0.42\linewidth}
\centering
\includegraphics[width=\columnwidth]{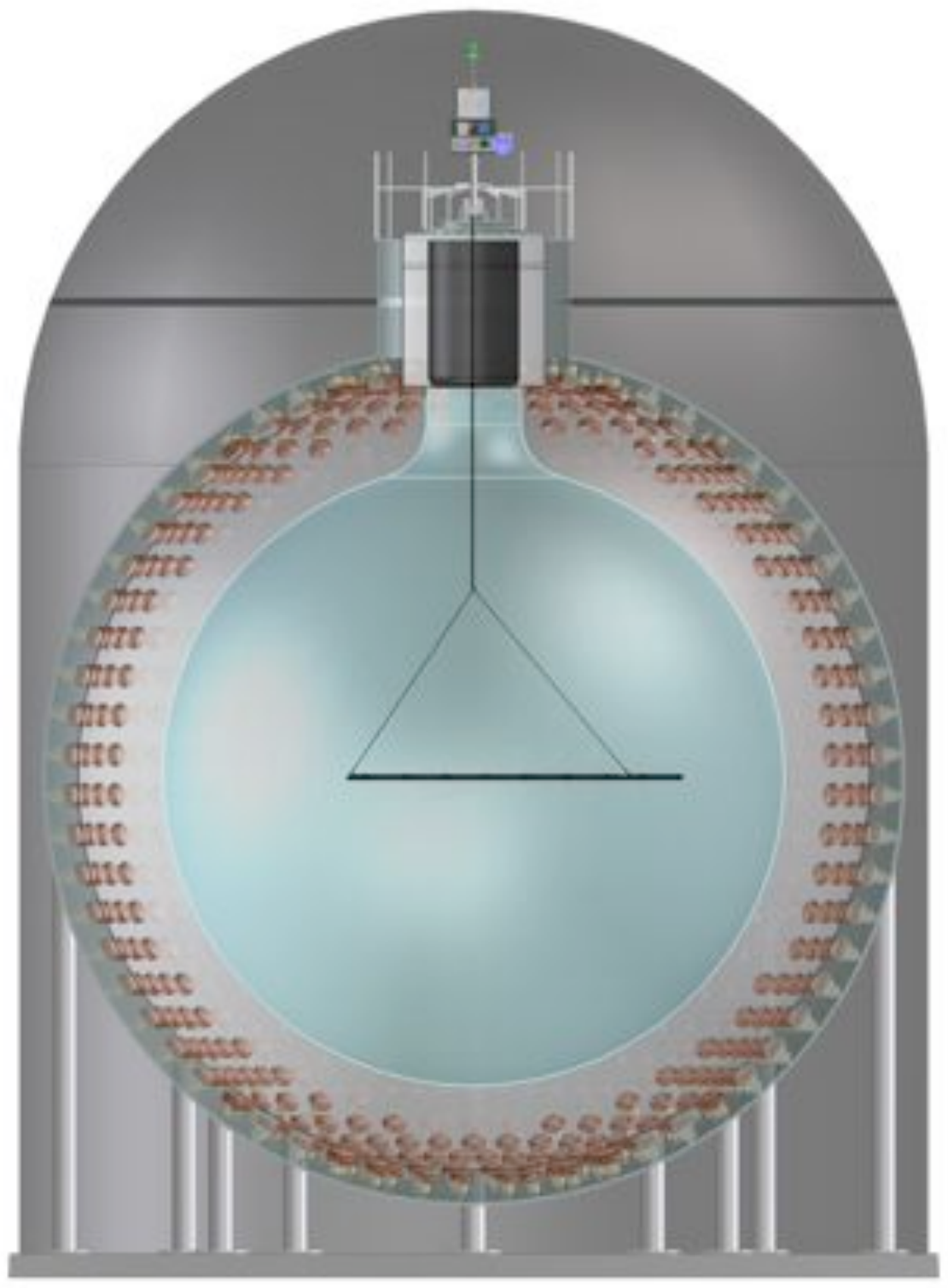}
\caption{Illustration of the calibration  system in the KamLAND detector: 
A radioactive source was attached to one end of the pole.  It was positioned throughout the fiducial volume by adjusting the orientation and length of
      the pole.   Additional $^{60}$Co pin sources, used for
monitoring the pole position, were located along the pole. 
  Two cables and a spool system manipulated the pole position and provided electrical connections to
      instrumentation in the pole. Access to the detector and manipulation of the system was
      through a glovebox on top of the chimney. 
 }
\label{fig:concept}
\end{minipage}
\hspace{0.4cm}
\begin{minipage}[b]{0.52\linewidth}
\centering
\includegraphics[width=80mm, clip, trim= 15mm 22mm 5mm 10mm]{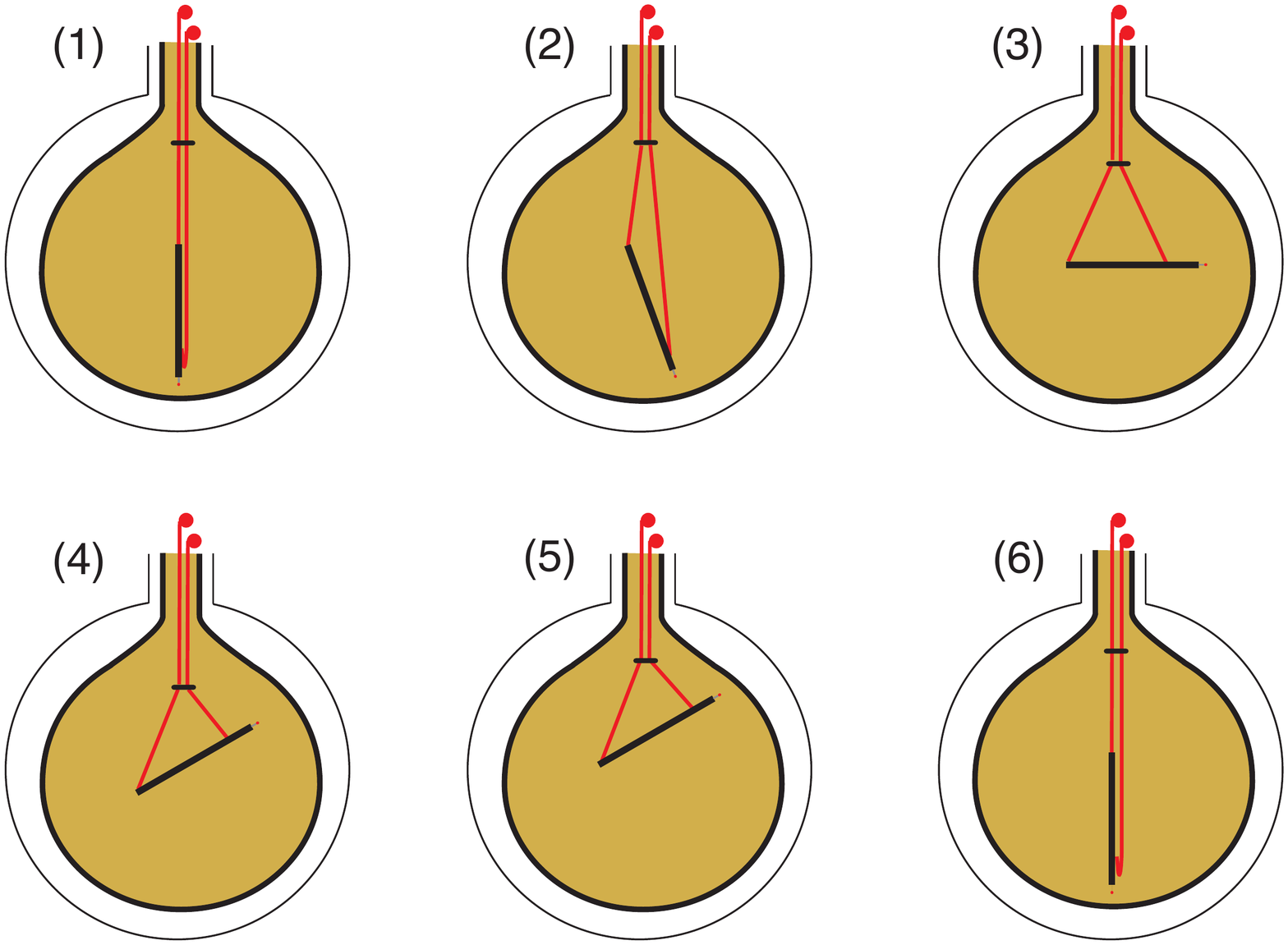}
\vskip 1.25cm
\caption{A typical deployment sequence for off-axis calibration: (1) During insertion into the detector, the pole was vertically suspended from the far cable with some slack in the near cable. (2) Once the pole was inside the detector, the near cable was raised, which removed the slack and pulled the pole off axis. (3) The far cable was then lowered and the near cable was raised to move the pole to the horizontal. (4) The near cable was shortened to bring the pole above the horizontal. (5) The far and near cables were raised simultaneously to bring the source closer to the balloon. (6) To prepare the pole for retraction, the near cable was lowered to return the pole to vertical.}
\label{fig:FourPiDeployment}
\end{minipage}
\end{figure}

The calibration system, depicted in Fig.~\ref{fig:concept}, is described in detail in section~\ref{sec:design}.  The design achieved access to the full fiducial volume of the detector while meeting strict safety challenges, primarily maintaining the integrity of the thin nylon/EVOH balloon and preserving the high radiopurity of the LS (section~\ref{sec:operations}).    The design was further challenged by the fact that all access to the fiducial volume had to be achieved via a 15.24~cm diameter flange that connects the glovebox to the detector's chimney region.   The calibration was done by deploying radioactive sources to mechanically determinable positions throughout the fiducial volume, especially near the boundary of the fiducial volume.  The reconstructed positions of the sources were then compared to these known positions.   Systematic biases in reconstructed position were thus measured and the fiducial volume systematic uncertainty established.    Reproducible and accurate mechanical positioning was therefore a critical design requirement.    The calibration data were also used to measure biases in the event energy reconstruction. Section~\ref{sec:performance} contains a discussion of the data analysis procedure.

\section{Calibration System Design}
\label{sec:design}

\subsection{System Overview}
\label{sec:overview}

The calibration system consisted of a segmented calibration pole, a variety of radioactive sources, two control cables for the manipulation of the pole inside the detector, and a motion control and cable spool system inside a glovebox on top of the detector. Monitoring instrumentation was integrated into the calibration pole, and read-out lines were embedded in the control cables.  

\begin{figure*}[th]
\begin{center}
\includegraphics[scale=0.55, clip, trim= 0mm 100mm 0mm 20mm]{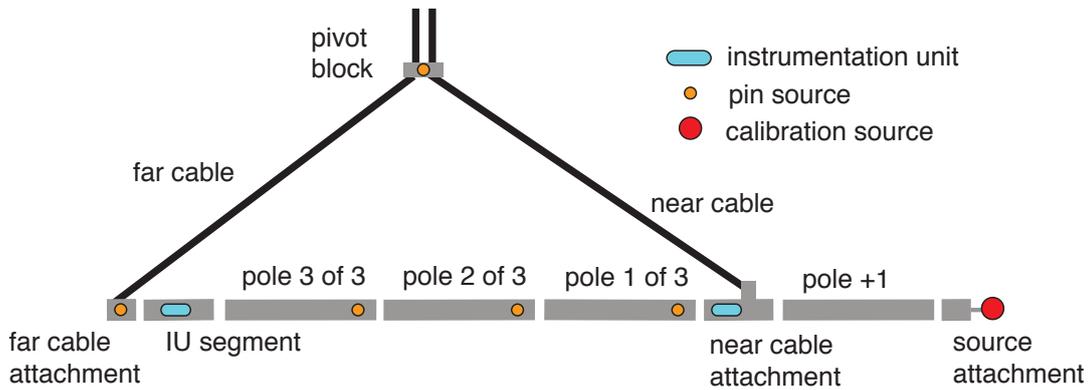}
\caption{Schematic diagram of the main components of the calibration pole:    The main body of the pole was composed of 90~cm long pole segments.  Short segments attached to the suspending cables and contained instrumentation.  The source attached to the tip of the pole+1 segment one meter beyond the near cable attachment. This figure shows the ``3+1'' pole configuration with three pole segments between the cables and one beyond the near cable. }
\label{fig:namingDiagram}
\end{center}
\end{figure*}

Fig.~\ref{fig:FourPiDeployment} illustrates a typical deployment sequence. Calibration at positions off the axis of symmetry of the detector (off-axis mode) required the deployment of the segmented calibration pole with two control cables. During insertion and retraction of the system, the pole, attached to both cables, was passed through the narrow access flange and chimney region. The pole was lowered vertically with all weight borne by the cable that was attached furthest from the source-end (the far cable).  The calibration pole moved into off-axis position when slack was removed from the cable attached near the source end (the near cable). The near and far cables are labeled in Fig.~\ref{fig:namingDiagram}. Adjustment of the relative lengths of the cables controlled the zenith angle of the calibration pole. To expand the radial reach in the upper and lower regions of the detector, the entire configuration was translated vertically.  The azimuthal position was varied by rotating the entire glovebox, which was mounted on a rotary stage, prior to the calibration deployment.  

The calibration pole consisted of several 90~cm long hollow titanium pole segments to which a radioactive calibration source was attached at one end. The number of segments suspended between the cable attachment points varied from one to six, and there was an additional segment to offset the source.  This is denoted as an ``N+1'' pole configuration. To increase the radial reach of the source, the segment furthest from the source-end could be replaced with a weighted segment, which consisted of a titanium pole segment containing a stainless-steel weight. With the weighted segment, the center of gravity of the pole was shifted away from the source end. A configuration with a weighted segment is denoted as ``N+1W''.  See Fig.~\ref{fig:namingDiagram} for a diagram of the calibration components.   

A ``pivot block'' held the two cables together above the pole, forming a triangle from the two lower portions of the cables and the pole.  It acted as a guide to the motion of the cables, preventing them from touching the nylon/EVOH balloon in the region where it tapered into the chimney, and improved the maneuverability of the pole system.

Radioactive calibration sources were attached to the end of the calibration pole with a special source attachment segment. This segment provided a standardized connection for a variety of radioactive sources.  In addition to the calibration source at the end of the pole, the system used up to eight additional $^{60}$Co pin sources for monitoring the pole position. They were inserted at intervals along the length of the pole and in the pivot block.  These embedded sources provided a means to perform high-accuracy relative position measurements because the relative distances between the pin sources were known to a few millimeters.  This served as a better constraint than the absolute position of the pole within the detector.  
  
During normal data taking, the inner detector\footnote{The inner detector refers to the volume enclosed by the 18~m diameter stainless-steel sphere.} is hermetically sealed.  The point of entry into the active volume is through an access flange at the top of the detector chimney.   Access is controlled by two gate valves in the chimney which were opened to perform calibration deployments. The 15.24~cm access flange and gate valve are the limiting apertures for insertion of the calibration pole and other objects into the inner detector.  


A glovebox, mounted on top of the chimney, housed the motion control spool system, the two control cables, the calibration pole segments, and other specialized hardware for the deployment system. The calibration pole was assembled and deployed from inside this glovebox by a team of three operators. The glovebox provided an ultra-clean environment with a controlled atmosphere from liquid nitrogen boil-off. Calibration sources were stored outside the glovebox in sealed storage bags and brought through a transfer box into the glovebox for deployment. 

The calibration system could also be utilized in a simpler mode of operation, the on-axis mode. In the on-axis mode, sources were deployed to positions along the central vertical axis of the detector. The radioactive source was attached directly to one of the control cables, without any pole segments, and lowered along that axis.  The precision of the source positioning was 2~mm.  The on-axis mode required significantly less manpower to operate and minimal pre-deployment assembly.  On-axis calibrations were performed typically once every two weeks to monitor changes in PMT gain and scintillator properties.  In the off-axis mode, calibration sources were deployed throughout the fiducial volume, especially near the boundary. The remainder of this paper elaborates fully on the operations of the calibration system in the off-axis mode.

\subsection{Deployment Hardware}
\label{sec:hardware}

%
\begin{figure*}[th]
\begin{center}
\includegraphics[width=0.9\linewidth]{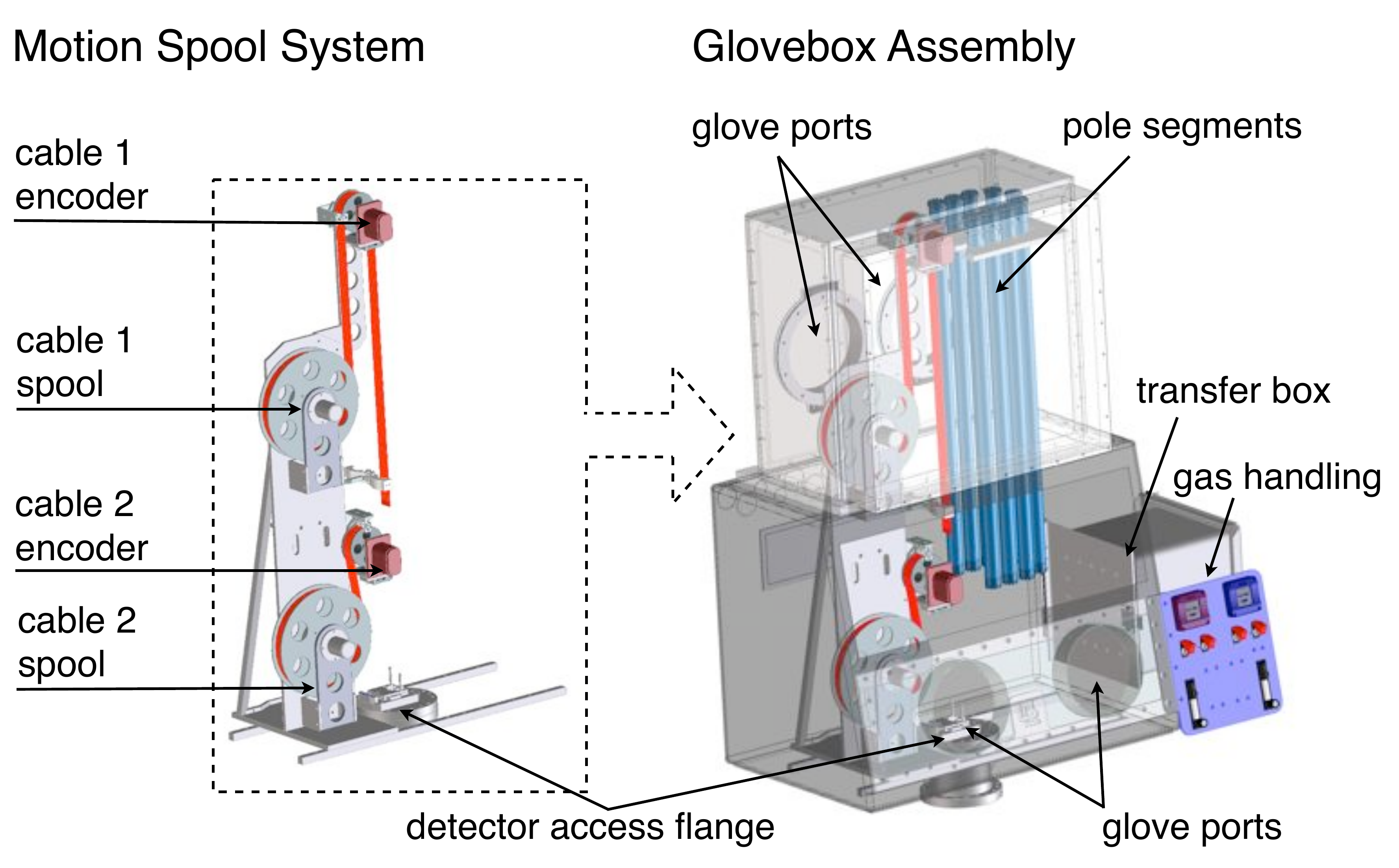}
\caption{Diagram of the motion spool system (left) and the glovebox assembly (right): The motion spool system was used inside the glovebox to manipulate the two control cables. The glovebox assembly was mounted on a rotary stage that controlled the azimuthal ($\phi$) rotation of the calibration system and was connected to the chimney of the detector through the access flange.}
\label{fig:FourPiGloveBoxDiagram}
\end{center}
\end{figure*}
Central to the deployment of the calibration system was the glovebox assembly, shown in Fig.~\ref{fig:FourPiGloveBoxDiagram}. 
To allow for the storage and vertical assembly of the 90~cm long pole segments, an acrylic extension to the main glovebox was constructed. The glovebox extension had a pair of glove ports.  This allowed a second operator to work in the space directly above the main glovebox volume.  With the glovebox extension, the longest pole segment that could be manipulated in the glovebox was 90~cm. 

All tools and hardware inside the glovebox were tethered to prevent accidental loss of items into the inner region of the detector.  Aside from being irretrievably lost, anything dropped into the inner detector could catastrophically damage the balloon or create permanent sources of background radiation.  The safe assembly and lowering of the calibration pole into the detector posed a unique design challenge for the calibration system. Specialized flange covers, along with secondary safety features in the segmented calibration pole, were implemented to ensure that no hardware could be accidentally lost in the detector.  
 
\begin{figure}[thpb]
\begin{minipage}[b]{0.42\linewidth}
\centering
\includegraphics[width=\columnwidth]{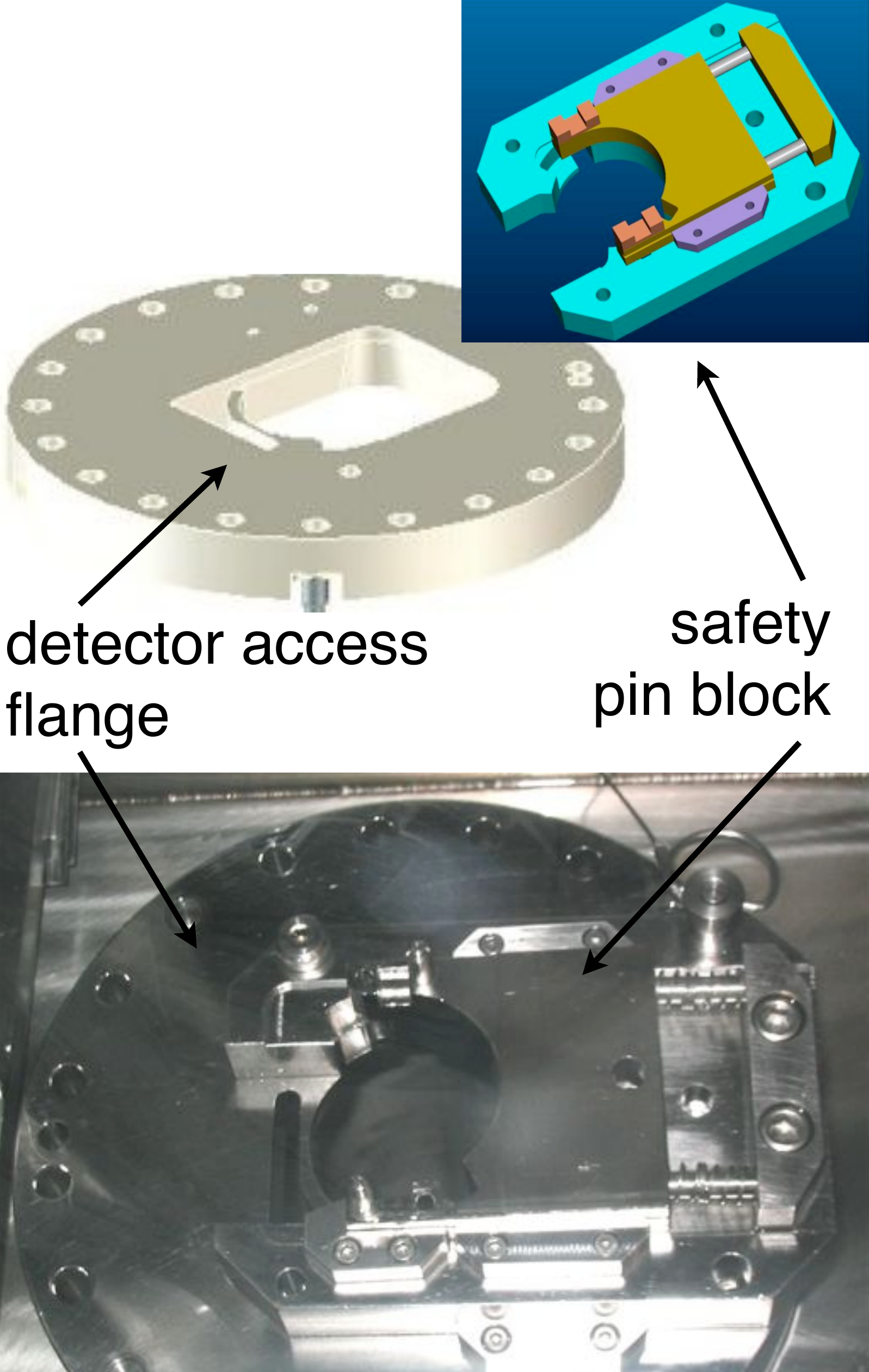}
\caption{Engineering model and photograph of the access flange with safety pin block: A specially designed access flange allowed the assembly, insertion, and retraction of the calibration pole while preventing any parts of the system from falling into the detector. The pole's safety pins engaged in the pin block and provided a two-step locking mechanism.}
\label{fig:pinblock_drawing}
\end{minipage}
\hspace{0.5cm}
\begin{minipage}[b]{0.54\linewidth}
\centering
\includegraphics[width=\columnwidth]{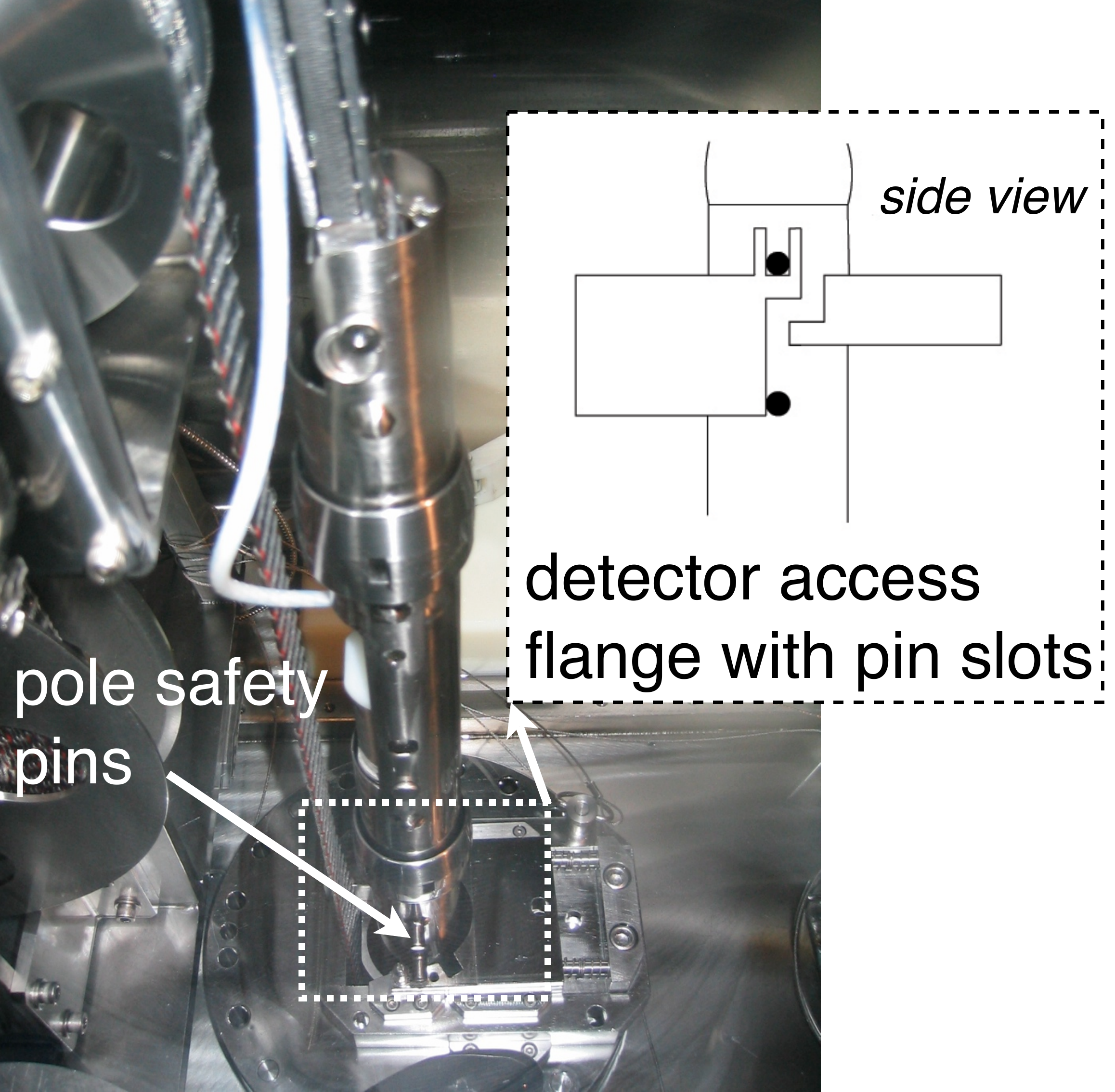}
\vskip 1.0cm
\caption{Photograph of the access flange with a calibration pole resting in it before deployment: The schematic insert illustrates the mechanism of the pole's safety pins and the pin block. The black dots represent pins, which were welded to the side of the pole segment.  These pins rested in the pinblock, preventing the pole from moving through the flange unless it was lifted and maneuvered through a series of small turns.}
\label{fig:pinblock}
\end{minipage}
\end{figure}

The 15.24~cm access flange at the bottom of the glovebox served as the interface between the glovebox volume and the chimney region of the detector.  A special flange cover, the pin block, provided a barrier for the pole segments during the assembly and dis-assembly phase; these components are shown in Fig.~\ref{fig:pinblock_drawing}.  Each pole had two sets of pins perpendicular to the pole segment that engaged in the pin block.  These pins rested in the pin block during the assembly and dis-assembly of the pole and provided a secure position for each individual segment.  With a pole sitting in the pin block, there was no free path from the glovebox into the detector.  Once the next pole segment was attached, the entire assembly was lifted so that the upper part of the pin block could be swung out of the way.  It was also possible to lower the assembly down through the pin block through a series of deliberate turns, as illistrated in~Fig.~\ref{fig:pinblock}.


The calibration pole was attached to two control cables that provided mechanical support and control.  A custom 2.54~cm flat woven cable was designed to meet the load, electrical, and stability requirements of the system. The flat design of the cables provided additional rotational stability compared to a conventional twisted cable design.  
The cables, manufactured by Woven Electronics, were formed from eight stainless-steel cables, for strength, and woven together with nylon. Included in the nylon weave were seven Teflon-coated, 30 AWG (American weight gauge) copper wires. They were used to make electrical connections to the instrumentation units (\ius) in the pole.  The cables were marked every meter with a series of stainless-steel staples; the pattern of staples within each mark indicated the distance along the cable.  The distances between successive cable marks were known to millimeter precision. 
%


\begin{figure}[th]
\begin{minipage}[b]{0.47\linewidth}
\centering
\includegraphics[width=\columnwidth]{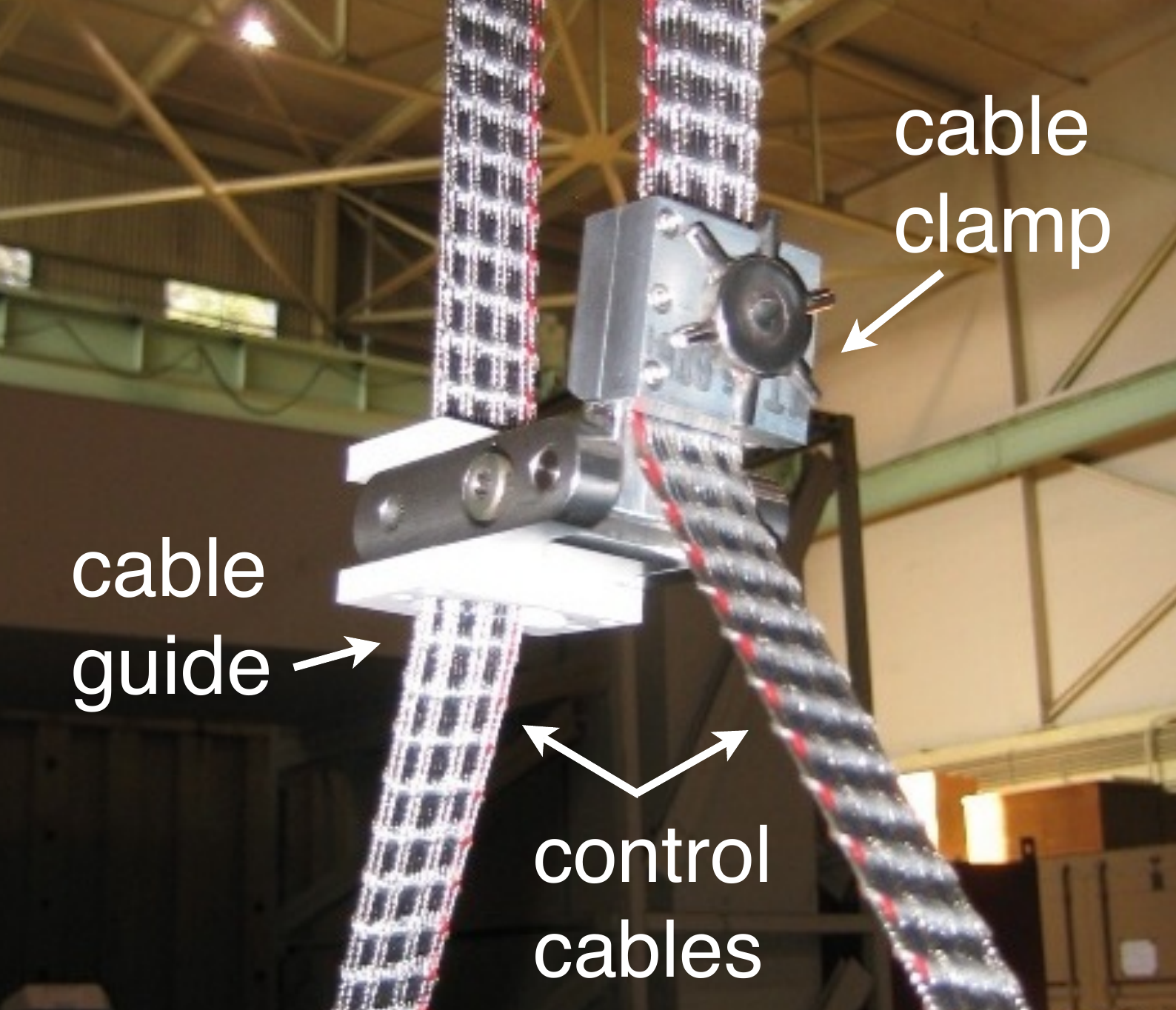} 
\includegraphics[width=\columnwidth]{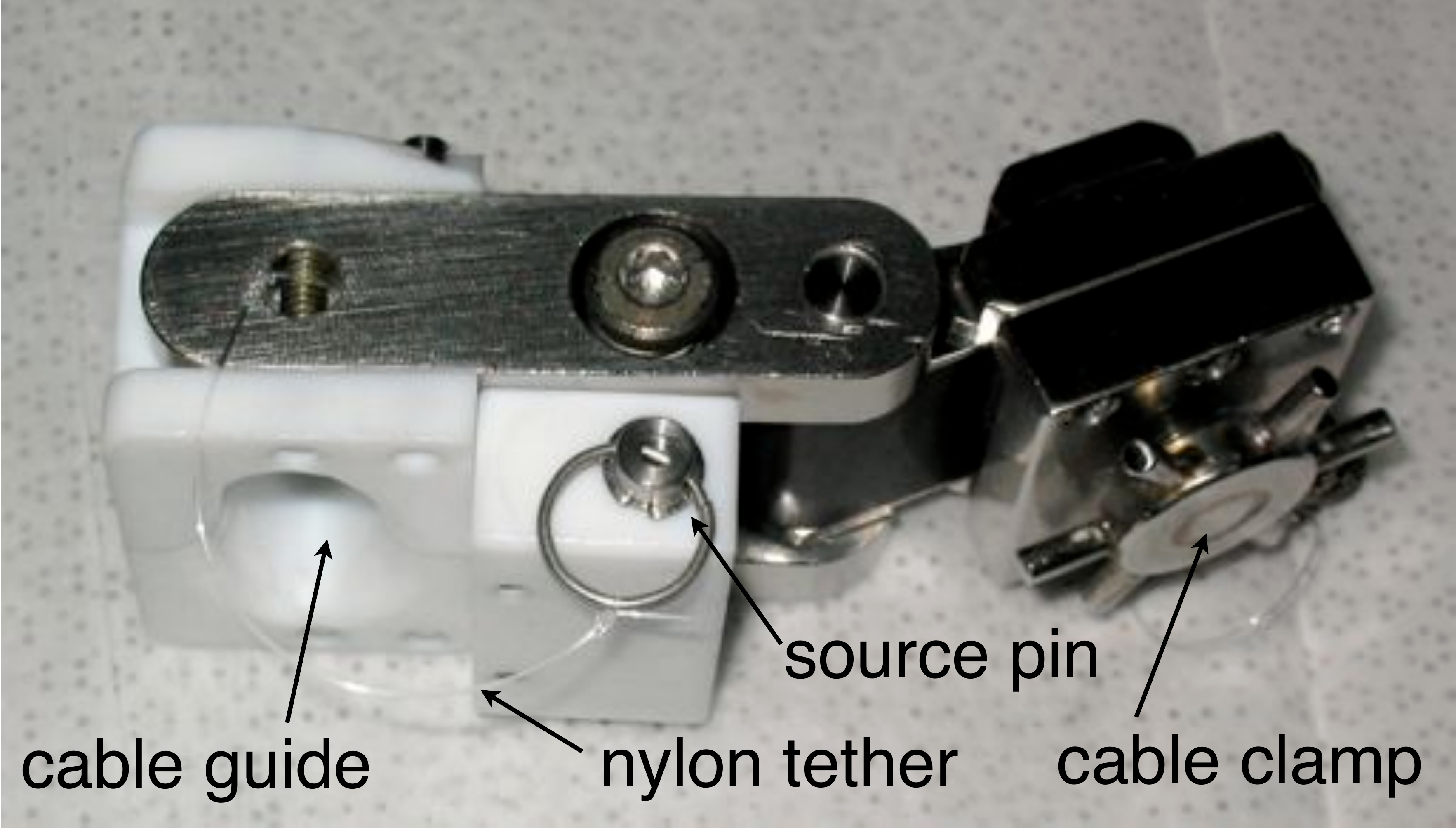}
\caption{Pivot block: The pivot block was clamped to the far cable (on the right), which supported the non-source end of the pole.  The near cable (on the left) slid freely through the Teflon guide, which was held at a right angle by the lateral forces from the cables.  The bottom photo is a labeled close-up. The source pin contained a $^{60}$Co source for monitoring the position of the pivot block within the detector.}
\label{fig:pivblock}
\end{minipage}
\hspace{0.5cm}
\begin{minipage}[b]{0.47\linewidth}
\centering\includegraphics[width=\columnwidth]{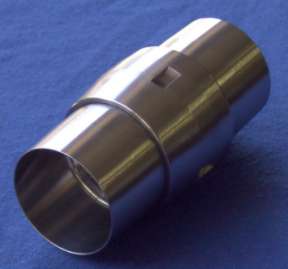}
\includegraphics[width=\columnwidth]{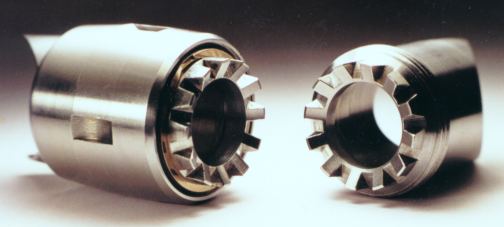}
\vskip 2.1cm
\caption{Bicycle Torque Couplings (BTC) in closed (top) and open (bottom) position: BTCs were used for coupling the individual calibration pole segments.  The photographs are courtesy of the manufacturer, S and S Machine \cite{BTC}.}
\label{fig:btc}
\end{minipage}
\end{figure}

A critical component of the off-axis system was the pivot block, which is shown in Fig.~\ref{fig:pivblock}.  It defined the geometry of the control cables and ensured that they passed through the access flange and chimney without touching the thin nylon/EVOH balloon. The pivot block was fixed to the far cable at a position that defined a stable, triangular configuration with the pole.  The chosen position kept the forces on the cables and the buckling stress on the calibration pole acceptably small. The pivot block was fixed to the far cable by two Viton covered stainless-steel plates that clamped to the woven cable.  A locking pin prevented the pivot block from falling off the cable in the event of mechanical failure. The near cable moved freely through the other side of the pivot block, which consisted of an oval shaped Teflon-lined guide.
%

\begin{figure}[tpb]
\begin{minipage}[b]{0.44\linewidth}
\centering
\includegraphics[width=\columnwidth]{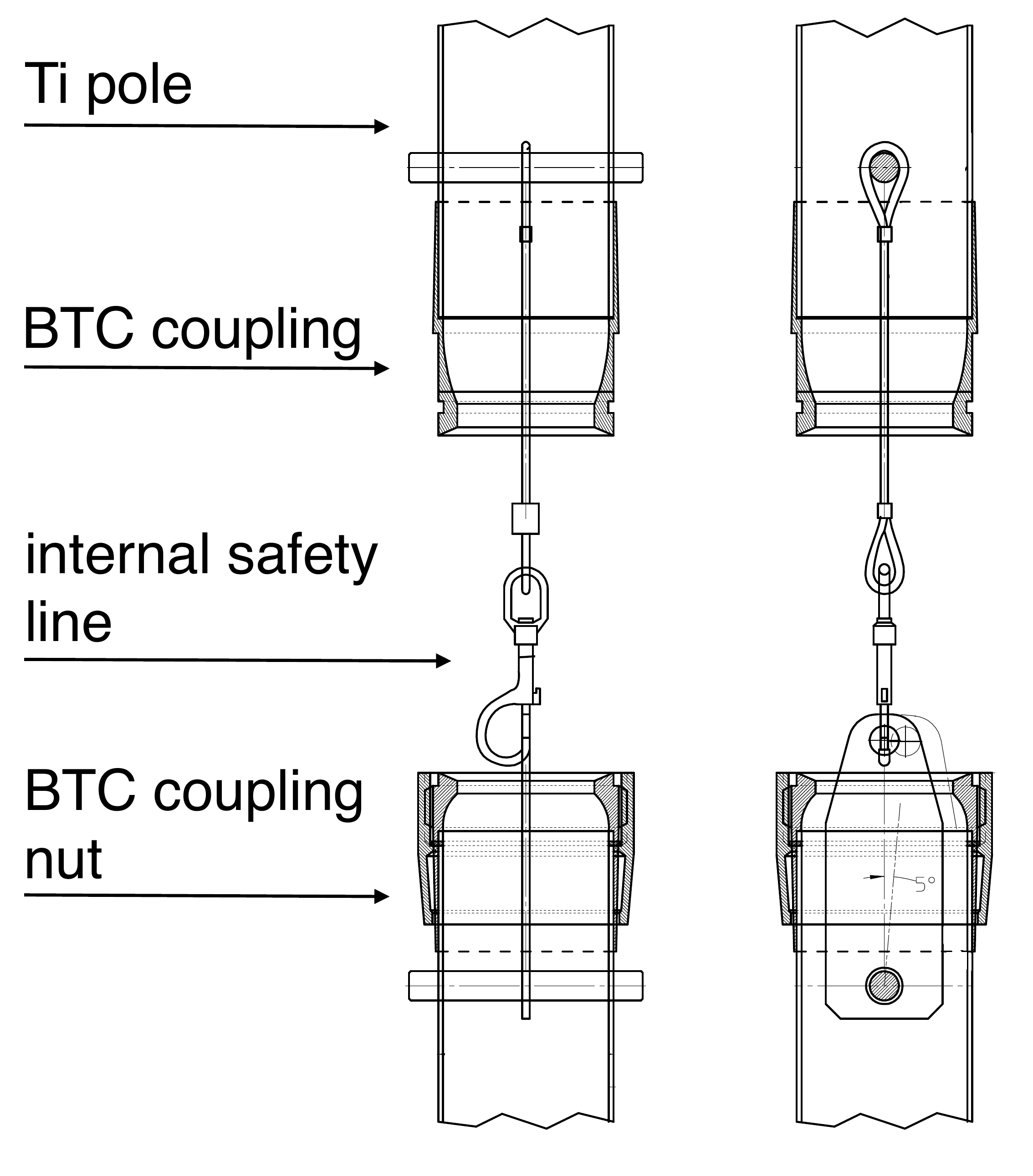}
\vskip 1.5cm
\caption{Engineering drawing of the internal safety lines between the calibration pole segments: The internal safety lines provided a secondary fail-safe mechanism for the calibration pole assembly and ensured that no parts of the system would be lost inside the detector.}
\label{fig:fourPiPoleSafetyLine}
\end{minipage}
\hspace{0.5cm}
\begin{minipage}[b]{0.52\linewidth}
\centering
    \includegraphics[width=\columnwidth]{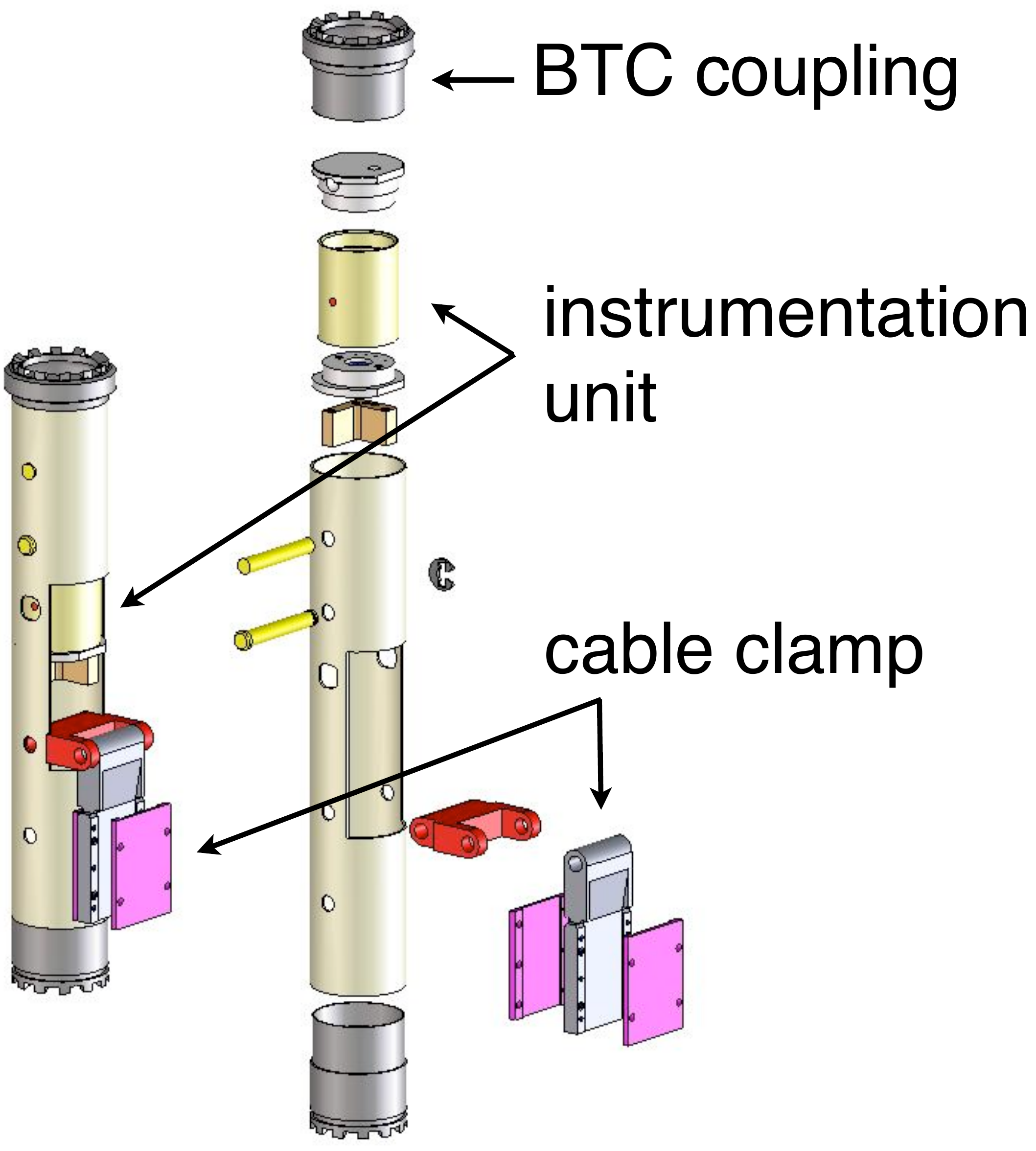}
\caption{A model of the far control cable attachment to the calibration pole: The cable was connected to the pole through a friction-based cable clamp with Viton inserts. An ~\iu\ was integrated into the pole and read out through the control cable. BTC couplings connected the cable attachment segment to the other pole segments.}
\label{fig:cable_attachment}
\end{minipage}
\end{figure}

The pole segments were constructed out of titanium tubing with a 3.81~cm outer diameter and a 0.9~mm wall thickness.  Titanium was chosen to minimize the weight of the assembled pole while maintaining compatibility with the LS. The weight of the longest calibration pole, in the weighted configuration, was 9.6~kg. Titanium bicycle torque couplings (BTCs), by S and S Machine, were welded to the ends of each pole segment to make the connections between the pole segments, as shown in Fig.~\ref{fig:btc}.  The pole segments had an internal stainless-steel wire tether to serve as a secondary restraint, which is seen in Fig.~\ref{fig:fourPiPoleSafetyLine}.

%

\begin{figure}[tpb]
\begin{center}
\includegraphics[width=0.6\columnwidth]{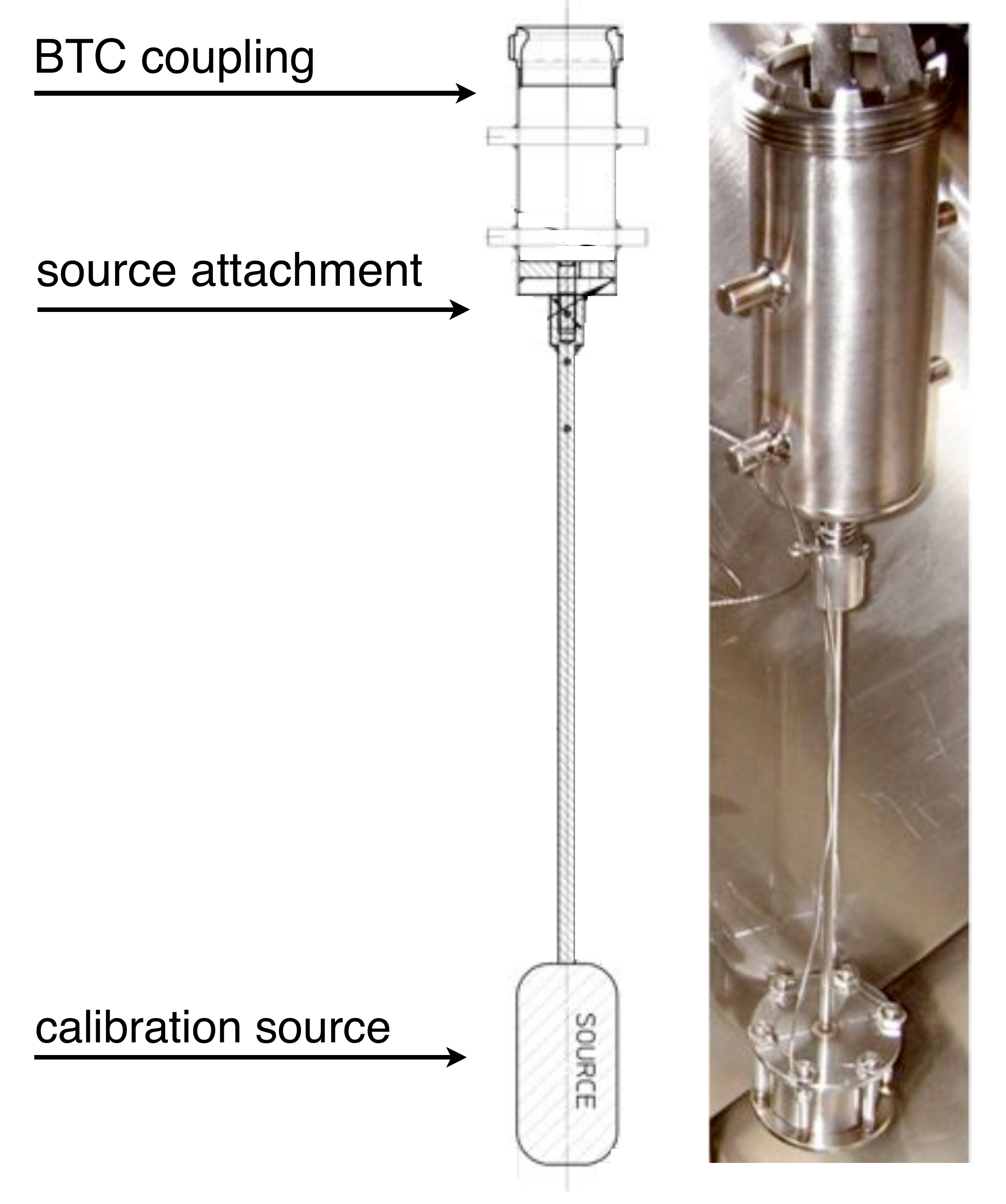}
\caption{A photograph and a drawing of the specialized source attachment with a calibration source: Depicted is a neutron calibration source (AmBe) attached to the source attachment segment. The source attachment couples via a BTC to the calibration pole.  Safety pins on the source attachment segment coupled to the source to prevent accidental loss during the deployment.}
\label{fig:source_attachment}
\end{center}
\end{figure}

Each cable was attached to a pole segment with a two-sided cable clamp that was lined with Viton. The Viton provided more friction on the composite, woven cable and prevented wear on the cable from  the stainless-steel plates. The flat cable was wound around the core of the cable clamp for increased friction and load strength. The cable clamps were mechanically linked to the calibration pole with hinges.  They were tested during prototyping and commissioning with a 12~kg load on both dry cables and cables that had been immersed in LS. The tests confirmed that the clamps could hold the weight of the longest pole configuration. 

%

The cable clamps attached to special short pole segments, shown in Fig.~\ref{fig:cable_attachment}, by means of a hinge mechanism that could align itself with the pole during deployment and retraction. Both cables connected to \ius\ that were integrated into the calibration pole close to the cable attachment points. At the near cable end of the pole, the ~\iu\ was integrated into the attachment section.  For the far cable end, a specialized pole segment held an ~\iu\ at the cable attachment point.

%

Calibration sources were connected to a special short attachment segment that coupled, via a BTC, to the pole on one side and connected with a thread and safety pin to the source on the other side.  This segment mirrored the connection hardware on the on-axis calibration system so that all of the sources constructed for that system could be deployed with the off-axis system.  The source attached to a threaded rod as shown in Fig.~\ref{fig:source_attachment}.  Locking pins were employed as a secondary safety mechanism to prevent the source from accidentally unscrewing.
%

\begin{figure}[htpb]
\begin{center}
\includegraphics[width=0.6\columnwidth]{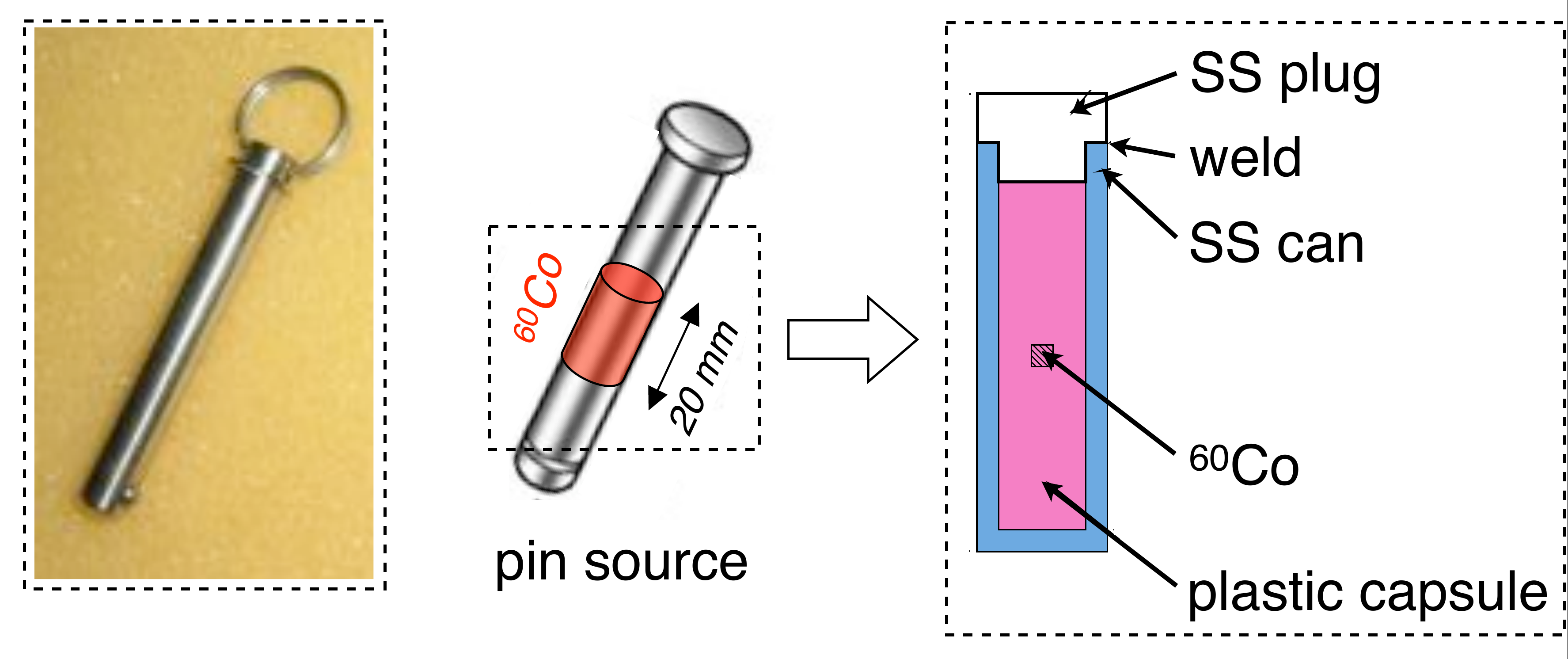} 
\caption{A photograph and a drawing of the doubly-encapsulated $^{60}$Co pin sources used for monitoring the position of the calibration pole: Sealed $^{60}$Co sources were located inside stainless-steel pins.  They were used to monitor the position of the individual calibration pole segments and the pivot block.}
\label{fig:pinsource}
\end{center}
\end{figure}

Fig.~\ref{fig:pinsource} shows a photograph and drawing of the $^{60}$Co pin sources used for relative position measurement. The sources had activities ranging from 100 to 250~Bq; they were fabricated as sealed sources by AEA Technology QSA and embedded inside stainless-steel pins.  The pins had a ball-and-spring lock which allowed them to be inserted into designated slots in each pole segment and into the pivot block.  


\subsection{Motion Control System}
\label{sec:controls}

The position of the calibration system was manipulated by adjusting the lengths of the two control cables.  Each cable was stored on a rotating spool that was turned by a brushless servo motor, Parker Compumotor SM232AE, with integrated shaft pulley encoder and fail-safe brake, through a 100:1 planetary gearbox, Parker Bayside RS90-100.  Electrical power to each motor was supplied through a Parker Compumotor Gemini GV-U2E drive unit, which used feedback from the motor's integrated encoder to allow its rotational speed and acceleration to be determined through digital step-and-direction signals.  The drive unit also enforced a motor torque limit that was set to be marginally higher than the minimum necessary to accelerate the weight of the system.  This limit minimized the danger of damage to the detector in case of unexpected mechanical interference. The motors, gears, and encoders of the motion spool and pulley system were housed in leak-tight stainless-steel enclosures to avoid contact with the LS vapors in the glovebox. For all cables and materials exposed to the LS and its vapors in the glovebox, special considerations for material compatibility and cleanliness applied, described in Section~\ref{sec:radio}. 

The relevant step commands were generated by a Parker Compumotor 6K4 four-axis controller, under the direction of control software that monitored the unspooled length of the cables.  These length measurements were relative; changes in the cable length were monitored by the passage of the cable over encoder pulleys mounted in the glovebox between the spool and the glovebox flange.  The active element of the pulley was a US Digital S1-1024 optical encoder component.  The quadrature pulses from this device were counted by the 6K4 motor controller.  There was not a simple proportionality between the motor's rotational speed and the linear speed of the cable because the effective radius of the spool changed during deployment.  Consequently, the system could not rely on the 6K4 controller's internal feedback and scaling functions to coordinate the motor with the encoder measurement. Instead, appropriate motor rotation commands were generated by control software on a PC.  When motion was desired, the computer calculated the motor rotation needed to move 80\% of the way to the target length and issued a command to move through that distance.  This process was iterated until the length fell within the specified tolerance of 1~mm from the target value.

The software was tiered, with a user interface layer that functioned independently of the hardware control layer.  The two layers communicated through the Java Remote Method Invocation protocol; this communication was limited to reading or setting the values of named parameters, and registering for notification when a named parameter changed.  There was, for example, a parameter that represented the current length of each cable. The values of all the parameters of the system were logged to a MySQL database as they were modified, together with periodic baseline logs of all parameters.

The software-based control system was used only after the pole had been assembled and moved to its starting position, where the initial cable lengths were known. An operator panel, Parker Compumotor RP240, was used to maneuver the system under manual control during the pole assembly and dis-assembly procedures.

Electrical power distribution was arranged to minimize pickup of the high-current motor drive by sensitive electronics, most notably the \ius.  An isolation transformer was used to separate the motor drive and 6K4 controller power from the other systems; an optically isolated USB hub, B\&B Electronics UISOHUB4,  was inserted between the control PC and the motor controller in order to maintain the separation of the ground lines, which was measured to be greater than 50~M$\Omega$.




\section{Radioassay \& Material Compatibility}
\label{sec:radio}

KamLAND \ls\ is composed of 20\% pseudocumene, 80\% dodecane and 1.36 g/l of PPO \cite{KLprl3}. The \ls\ was purified during the construction of KamLAND to reach the low levels of $^{238}$U,   $^{232}$Th and $^{40}$K needed for sensitivity to \antinu s. During the initial data-taking stage of KamLAND, the contamination from these isotopes was determined to be $(3.5\pm0.5)\times10^{-18}$~g/g $^{238}$U~\cite{KLprl2},  $(5.2\pm0.8)\times10^{-17}$~g/g $^{232}$Th ~\cite{KLprl2} and $(1.90\pm0.04)\times10^{-16}$~g/g $^{40}$K. Work is now underway to further reduce these contamination levels~\cite{KishimotoTAUP2007} to increase KamLAND's sensitivity to $^{7}$Be solar neutrinos.  When deployed, the components of the calibration system were submerged in the \ls\ for up to 150 hours at a time. It was therefore imperative that the material radiopurity and compatibility of the calibration system be considered during construction.


\subsection{Material Compatibility and Cleaning Procedures}

Pseudocumene is an organic solvent that affects the structural integrity of many materials. Extensive tests on the compatibility of materials with \ls\ have been done by the KamLAND collaboration and others \cite{Compatibility1983}. Titanium, 304 and 316 stainless-steel, gold, nylon, Teflon  and Viton were all found to be chemically compatible with \ls. Certain types of acrylic appear to be compatible for periods of up to several years, but prolonged exposure of acrylic to pseudocumene can cause swelling, which exacerbates any weakness introduced during its processing. For this reason, cast acrylic and Lucite were chosen over extruded acrylic.  Machining of acrylic pieces was kept to a minimum to reduce internal stresses, which can lead to the formation of miniature stress fractures. Among all the materials in the calibration system design, titanium was the only material whose compatibility with the \ls\ had not been previously tested. A set of titanium BTC couplings were soaked in a sample of \ls\ for 9 months. Visual inspections following the soak indicated no deterioration of the BTC or discoloration of the LS, and the measured light attenuation length of the LS remained unchanged. 

The \ius\ and all sources were pressure tested to ensure their structural integrity.  The test consisted of pressure cycling the material four times to 275~kPa for 10~minutes and then maintaining this pressure for at least a 12~hour period.  After the test, materials were visually inspected.  The pressure test was performed prior to the certification procedure, which could further determine through gamma counting if sources leaked after the pressure test.

To minimize the introduction of radioactive contaminants, strict procedures were employed in the cleaning and certification of items entering the \ls\ volume.  All system components were UHV-cleaned and packaged in a class 10,000 clean room before shipment to the experimental site.  In general, upon arrival, items expected to come into direct contact with the \ls\ or with the calibration system's components underwent an additional cleaning procedure.  The procedures performed on site were done in a class 100 clean room.  To remove surface contamination, a procedure previously determined to be effective~\cite{CertificationProcedures} was adopted for all parts, whether they came into contact with the \ls\ directly or indirectly:
\begin{enumerate}
  \item Wipe down the material with a Kimwipe soaked in acetone and rinse with ASTM Type I de-ionized (DI) water. Wipe down again with isopropyl alcohol or ethanol and rinse with DI water.
  \item Put material in fresh DI water and place into an ultrasonic bath for 15 minutes.
  \item Remove material from DI water bath.  Immerse in strongest nitric acid (HNO$_{3}$) solution the material can withstand (0.1 - 1.0 M).  Soak for 5 minutes in an ultrasonic bath heated to 40$^\circ$~C.   Repeat twice.
  \item Place the material into a third ultrasonic HNO$_{3}$ bath for 1 hour.  
  \item Finally, remove the item from the HNO$_{3}$, rinse with DI water, and place the material into a container with clean \ls. Transfer the acid from this final acid bath into a counting bottle and count it in a germanium detector (section~\ref{sec:germaniumdet}) for certification.
  \item Leave the material in the \ls\ bath until certified.  Afterward, remove it and place it in a double-bagged, vacuum-sealed container.
\end{enumerate}
This general cleaning procedure was modified as needed for specific parts that were especially large or not chemically compatible with some of the solvents listed.


\subsection{Onsite Low Background Counting Facility}
\label{sec:germaniumdet}

The certification procedure used a p-type germanium detector with 28\% relative efficiency, custom-built from low-background materials by Canberra Inc., Belgium. The detector had previously been used for material certification in the Palo Verde experiment and during KamLAND construction.  The crystal holder and vacuum endcap were constructed from low activity aluminum, while all other cryostat parts were made from electrolytic copper. Gaskets near the crystal were fabricated from selected soft lead wire and screws and bolts were machined from pre-World War I steel. 

The detector was housed in a 5~cm thick copper inner shielding box that serves as the counting sample compartment. The inner shield was continuously flushed with boil-off nitrogen to displace airborne $^{222}$Rn.  The outer shield was formed from 20~cm of lead. The entire assembly was placed on a steel table to accommodate the vertical dipstick construction of the detector and to allow convenient sample insertion. Sections of the copper and lead shield can be moved by means of hydraulic cylinders for sample insertion. 

The counting efficiency was calibrated utilizing radon-loaded mine water and a potassium solution of known concentration. The validity of the efficiency and analysis program was verified by counting a sample of Table Mountain Latite rock for which good agreement was found in a cross-calibration by the Lawrence Berkeley Laboratory counting facility.

\subsection{Material Certification Procedures}
\label{sec:radioLimits}

\begin{figure}[ht]
\begin{center}
  \includegraphics[width=0.8\columnwidth, clip, trim= 10mm 10mm 20mm 10mm]{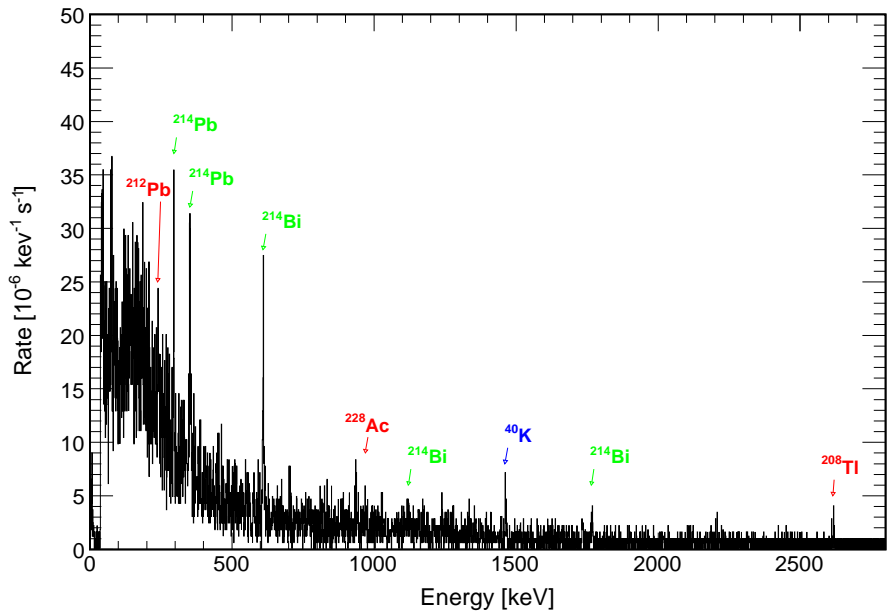}
\caption{Underground germanium detector background spectrum:  Here we indicate the gamma lines from internal $^{40}$K, $^{232}$Th and $^{238}$U and their daughters.}
\label{fig:GeBackgroundSpec}
\end{center}
\end{figure}

Liquid from each acid soak was counted with the low-background germanium detector.  Analysis of the detector background, shown in Fig.~\ref{fig:GeBackgroundSpec}, yielded detection limits of: 0.017~Bq  $^{238}$U, 0.015~Bq  $^{232}$Th, and 0.024~Bq  $^{40}$K, for a counting time of five days with our standard single bottle geometry at the face of the detector.  Items in direct contact with the \ls\ (Class I) were approved for deployment when a five-day count of the acid soak showed no activity exceeding a three-sigma significance after background subtraction.  To detect gross contamination of items that only came into contact with \ls\ vapors, or were not in direct contact with the deployment system (Class II), items were swiped with a pre-soaked alcohol wipe. These were counted for one day. An item was approved for installation if no activity above the background was observed.  The background subtracted activities obtained for each of the certified parts are listed in Table~\ref{tab:MaterialTable}.  These activities can be compared with the activity found in the nominal 5.5~m fiducial volume: 2.34 x 10$^{-5}$~Bq ($^{238}$U), 1.15 x 10$^{-4}$~Bq ($^{232}$Th) and 2.72 x 10$^{-2}$~Bq ($^{40}$K).  A conservative analysis assumes the measured component would saturate the entire fiducial volume upon deployment and the values are directly comparable.


Limited amounts of epoxy were used in the sealing of the \ius. We performed dry counts on three different types of epoxy resins which were candidates for use on the \ius.   The extra-fast setting epoxy, Hardman Double/Bubble Red, was found to be low in intrinsic activity and largely compatible with \ls. It exhibited some softening after extended soaking in \ls, especially when the epoxy was not mixed thoroughly. However, a test joint soaked for one week and stressed with a 4.5 kg hanging weight showed no deterioration in strength.  We list these results at the bottom of Table~\ref{tab:MaterialTable}.

The low limits for radioactive contamination prohibited the use of solder. All electrical connections inside the glovebox were made with Teflon-coated wires and gold-plated crimp connections. If connectors made from approved materials could not be procured, they were custom-made. Similarly, components made from metals other than those approved were gold-plated.  Vacuum-tight feed-throughs to the enclosures of the motor and encoder boxes were made by Ceramatec.


\subsection{In-Situ Radioactivity Measurements}
\label{sec:clean}

The KamLAND detector itself is a very sensitive low-background detector.  In order to characterize the intrinsic radioactivity of the off-axis system, we collected data from special deployments, which consisted of only the far cable deployed along the central axis of the detector.   We also looked for heightened levels of radioactivity following off-axis deployments.  We found no evidence for an increase in the bulk rate of events due to intrinsic background radiation.  Targeted searches for contamination from $^{40}$K and $^{208}$Tl also gave null results. 

\begin{figure*}[t]
\begin{center}
  \includegraphics*[width = \columnwidth,clip , trim= 10mm 10mm 10mm 80mm]{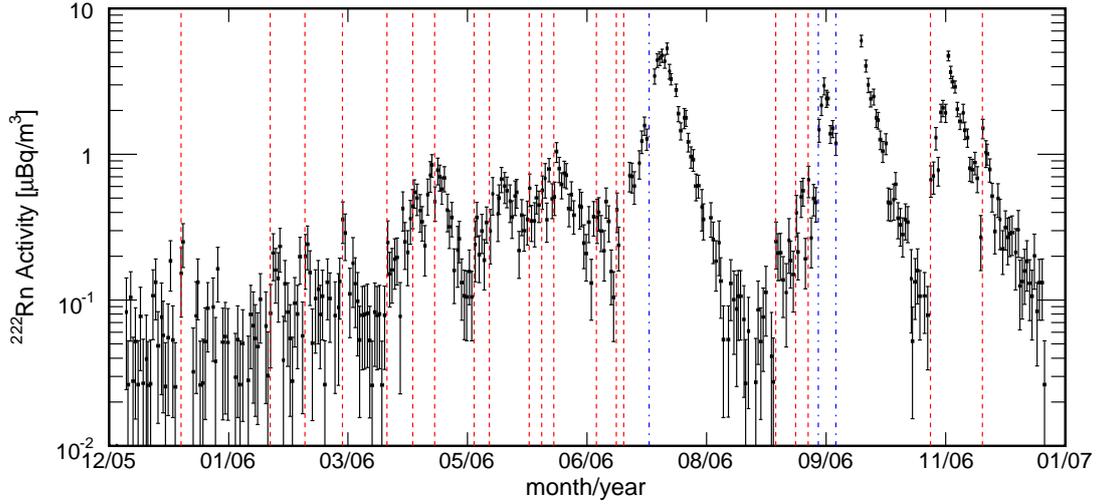}
\caption{One year of KamLAND data in which $^{214}$Bi-$^{214}$Po coincidences inside a 5.5 m fiducial volume are used to extract the $^{222}$Rn activity: The red (dashed) lines indicate the start of calibration periods in which we took data along the central z-axis of the detector.  The blue (dash-dot) lines indicate the beginning of off-axis calibrations.  Offsets in the decay curves from the calibration start time are due to long calibrations lasting up to a week. 
}
\label{fig:BiPoFit}
\end{center}
\end{figure*}

A more sensitive analysis using the short time correlation between $^{214}$Bi and $^{214}$Po decays indicated increased levels of \radon~following both on and off-axis deployments.  Fig.~\ref{fig:BiPoFit} shows the time dependence of the $^{222}$Rn rate.  This heightened level of \radon, and the subsequent decay from its daughters, is a negligible addition to the background for reactor~\antinu\ detection.   However the long-lived daughter of \radon, \lead, is problematic for the high-purity phase of KamLAND, which aims to detect the mono-energetic solar neutrinos from $^7$Be decay via elastic scattering off of electrons.  

Analysis of the cable-only deployment also revealed that small quantities of the \radon\ daughters appeared to have collected on the deployment cables while being stored inside the glovebox.  The daughters are thought to be introduced (via \radon) when the glovebox is opened to transfer materials and from \radon\ outgassing of materials in contact with the interior of the box.  Once deployed into the detector, the equilibrium with $^{222}$Rn is broken, and the daughters decay with the half-life of $^{214}$Pb down to the metastable state of \lead.  Owing to the short half-life of $^{214}$Pb, this background contribution was not detected in the data following off-axis deployments, and thus represented an additional contribution to the \lead\ activity.  

Additional studies of \radon\ contamination following on-axis deployments revealed that the level of post-deployment \radon\ would introduce 0.13 - 8.0 $\mu$Bq of \lead\ into the entire fiducial volume for a typical on-axis (one cable) deployment.  The exact amount depends on factors surrounding the handling of the deployment hardware and pre-deployment glovebox manipulations.  It was found that when proper handling and transfer occurred, the \radon\ introduction was in agreement with the measured amount of \radon\ in the nitrogen purge gas used to flush the glovebox.






%
%
\begin{table*}[!pt]
\vskip 3mm
\begin{threeparttable}
\begin{tabular*}{\columnwidth}{@{}p{1.8in} @{}p{0.7in} @{}p{0.7in} p{0.7in} p{0.8in} p{0.8in}@{}}
\hline
\hline
{\bf Item}         &{\bf Material}  & {\bf Method} &  {\bf $^{40}$K [mBq]}        &   {\bf $^{232}$Th [mBq]}  &  {\bf $^{238}$U [mBq]}   \\
\hline
\raggedright Titanium BTC (3AL25V)               & Ti       &    LS Soak &     -25 $\pm$   10 &      -3.8 $\pm$   3.1 &       5.4 $\pm$   2.1 \\

\raggedright Cable's Teflon Conductor            & T         &    LS Soak &     -15 $\pm$  13 &      -1.0 $\pm$   3.7 &       3.6 $\pm$   2.1 \\

\raggedright Cable's Nylon Mono-Filament         & N         &    LS Soak &     -18 $\pm$  14 &      -2.2 $\pm$   3.5 &       5.6 $\pm$   2.2 \\



\raggedright Near Cable                          & N, SS, T       &  Acid Soak &      311 $\pm$  141 &       33 $\pm$   27 &       9 $\pm$   14 \\

\raggedright Far Cable                           & N, SS, T    &  Acid Soak &      326 $\pm$  136 &       30 $\pm$   24 &       7 $\pm$   13 \\


\raggedright Titanium Pole Segment               &  SS, Ti          &  Acid Soak &      811 $\pm$  313 &      -8 $\pm$   42 &       45 $\pm$   25 \\

\raggedright Instrumentation Units               &  N, SS, V      &  Acid Soak &      38 $\pm$  21 &       1.2 $\pm$   3.5 &       0.9 $\pm$   1.8 \\

\raggedright Source Attachment, Cable Clamps,  and IU Holder &  SS, Ti      &  Acid Soak &      537 $\pm$  182 &      -10 $\pm$   23 &       19 $\pm$   14 \\

\raggedright IU Cables and Pins                  & G, N, T       &  Acid Soak &      436 $\pm$  223 &       34 $\pm$   32 &       7 $\pm$   19 \\

\raggedright Near Cable Attachment               & SS, Ti       &  Acid Soak &      571 $\pm$  238 &       38 $\pm$   27 &       10 $\pm$   20 \\

\raggedright Pivot Block                       & SS, Ti, V        &  Acid Soak &      195 $\pm$  138 &       94 $\pm$   41 &       3 $\pm$   17 \\



\raggedright Pivot Block Tool                   &  SS          &  Acid Soak &      435 $\pm$  181 &       38 $\pm$   34 &       34 $\pm$   21 \\


\raggedright $^{60}$Co Pin Sources  \tnote{1}            &  SS      &  Acid Soak &      407 $\pm$  235 &      -20 $\pm$   35 &       6 $\pm$   21 \\


\raggedright $^{203}$Hg Source \tnote{2}                 & SS       &  Acid Soak &      147 $\pm$  106 &       13 $\pm$   20 &       23 $\pm$   13 \\

\raggedright $^{210}$Po$^{13}$C Source  \tnote{3}        &  SS       &  Acid Soak &      128 $\pm$  61 &       3.3 $\pm$   8.7 &       9.0 $\pm$   5.8 \\


\raggedright Deployment Hardware                & N, SS, T      &      Swipe &      18 $\pm$  10 &       5.0 $\pm$   2.2 &       0.6 $\pm$   1.0 \\

\raggedright Tools and Containers               & SS       &      Swipe &      19 $\pm$  31 &      -4.7 $\pm$   6.1 &       5.4 $\pm$   3.6 \\


\hline
\hline
 & & & \bf{[mBq/g]}  & \bf{[mBq/g]} &  \bf{[mBq/g]} \\
\hline
\raggedright Hardman Double/Bubble Orange\tnote{4}             &  &  Dry Count &        252 $\pm$  56 &        156 $\pm$  16   &        124 $\pm$   8 \\

\raggedright Hardman Double/Bubble Red \tnote{4}                &  &  Dry Count &      46 $\pm$  24  &       5.3 $\pm$   3.9 &       4.0 $\pm$   2.1 \\

\raggedright Araldite 2011\tnote{4}                    &  &  Dry Count &      23 $\pm$  23 &       5.1 $\pm$   4.5 &      21 $\pm$   4 \\
\hline
\hline
\end{tabular*}
\vskip 3mm
\begin{tablenotes}
  \item[1] Fits to the 1173 and 1332 keV gamma lines of $^{60}$Co yielded 5.4 $\pm$ 5.3 mBq 
  \item[2] Fits to the 279 keV gamma line of $^{203}$Hg yielded 0.8 $\pm$ 2.4 mBq
  \item[3] Since there are no gammas associated with this source, liquid scintillation counting was utilized to determine a $^{210}$Po alpha rate resulting in a limit of 300 mBq
  \item[4] These are dry counts of the material.  The activity has been divided by the sample mass.
\end{tablenotes}
\end{threeparttable}
\caption[]{Tabulated results of the radioactivity measurements for materials and components used in the assembly and operation of the calibration system. The detection limits are described in  Section~\ref{sec:radioLimits}. Each sample rate has been scaled by the fraction of the counted liquid volume relative to the total soak volume. The material types are Gold (G), Nylon (N), Stainless-Steel (SS), Teflon (T), Titanium (Ti) and Viton (V).}
\label{tab:MaterialTable}
\end{table*}
\clearpage

\section{Operations}
\label{sec:operations}

Operation of the calibration system in off-axis mode required a team of several specialized operators.  A full deployment cycle, including pole assembly, deployment, data-taking, retraction, and dis-assembly, typically lasted from 1 to 5 days.   

Three operators were needed for both the assembly and dis-assembly of the calibration pole, with two of them working through the glove ports on the glovebox, shown in Fig.~\ref{fig:FourPiGloveBoxDiagram}.  A lower glovebox operator was responsible for leading the assembly and dis-assembly.  This operator made all the critical hardware connections: coupling the pole segments with the BTCs, attaching the control cables and readout, and positioning the pivot block.  An upper glovebox operator guided pole segments into place and assisted the lower operator.  A third person was needed for operating the manual drive controls and reviewing the assembly procedures.   The assembly and insertion into the detector typically took 2-3 hours depending on the pole configuration.  A similar time was required for the retraction and dis-assembly of the system.

Once the pole was deployed into the detector, operation required a system monitor and a control operator.  In principle, it was possible for one person to control the movement of the system.  However, a two-person rule was enforced to reduce the risk of operator error. During operation, the control operator entered pre-determined cable length values into the control program while the system monitor visually confirmed the correct movement of the cables through the window of the glovebox.  When the pole was in the off-axis configuration, the cables were moved at the speed of a few millimeters per second to preserve stability of the system.  Depending on the distance to be traveled, typical cable movements between pre-calculated off-axis positions took approximately 20 minutes.  Once the desired position was reached, motor brakes were engaged to lock the control cables in place.  The glovebox was made light-tight and then data were collected.

Typically, the system was operated over two shifts spanning 16-18 hours. If the same source was to be used on consecutive days, the pole was retracted until it could be secured in the pin block for the night.  This reduced the assembly time the next day and allowed for more data collection.


\subsection{Real-time Position Monitoring}
\label{sec:pm}

Real-time position monitoring was critical for safe deployment and operation. Maintaining a safe distance between the calibration pole and the delicate balloon was critical at all times.  During the operation of the calibration system, a set of redundant methods of protocol and instrumentation were used. 

The allowed safety region for the operation of the calibration pole consisted of a spherical volume centered within the detector and a cylinder leading up to the glovebox.  Real-time position monitoring was used to ensure that the calibration system did not move outside this safety region at any time.  The radius of the spherical safety volume was set in the control software to be 5.5~m.  This is significantly less than the 6.5~m radius of the balloon and provided a wide safety margin in the event of a positioning error.   

The primary position information was obtained from the motion control program.  The core function of the control software was to generate motion commands to achieve desired cable lengths in response to operator requests.  In addition, the control software had several other critical functions: it checked that it would be possible to transition to the requested cable lengths without moving any component of the system outside the designated safety zone, and it monitored the motion as it proceeded, ready to implement a forced stop if an unexpected condition arose. 

Visual observation of the cable motion and the reading of the cable marks served as the primary safety check during cable movements. Real-time data reconstruction was used to cross-check the system location inside the detector based on the reconstruction of the radioactive sources. The pressure and accelerometer readings of the \ius\ were designed to provide additional cross-checks.

To determine the pole position from the cable lengths, the online software calculated the potential energy 
\begin{equation}
V = m_p z_p + m_{near} z_{near} + m_{far} z_{far} ~.
\end{equation}
Here, $m_p$ is the mass of the pole assembly and $z_p$ is the vertical 
coordinate of its center of mass.  The mass of the portion of the near~(far) cable 
beneath the pivot block is $m_{near(far)}$, and the vertical coordinate of the 
center of mass of the cable is $z_{near(far)}$.  This potential energy function was 
numerically minimized as the angle of the pole together with the cable and pole lengths completely determined the configuration of the system.  Because the motion control program used the relative change in cable lengths to determine the system location, it was critical that the starting cable positions and calibration geometry were correctly input.  Once this requirement was met, the control program was capable of determining the position of the pole to several centimeters.

The control software verified, before motion began, that the requested step would be safely within
the allowed region even if one of the two motors were to fail at any point while 
the other carried out its commanded motion.   If this condition was not 
satisfied, then the requested step was automatically divided into an appropriate
number of shorter substeps.  This function became most relevant as
the system was raised or lowered vertically through the narrow zone between the 
detector and the glovebox, where substeps of 10~cm were typical.
Finally, the integrated number of motor revolutions was continuously
compared with the deployed cable length as measured by the encoders.  Motion
was stopped if there was a discrepancy larger than 60~cm.

During insertion and removal of the pole, the allowed safety zone in the chimney region was a small, 15~cm diameter cylinder.  Once the pole was fully inserted into the LS region, it was necessary to increase the radius of the cylindrical safety zone in order to move the pole away from the vertical. When the pole was hanging vertically by the far cable, the amount of slack in the near cable was constrained to be greater than 20~cm.


Because the control software provided an important redundant monitoring function, it was itself monitored by a watchdog circuit that would detect a sudden failure of the computer or software.  The control program toggled the state of a TTL output register on each pass through its monitoring cycle.  This signal was connected to a custom circuit, built around the Maxim MAX693CPE supervisor IC, that drove the interlock input to the motor controller.  Motor power was inhibited if no pulses from the computer were received for 1.6~s. Switches mounted on the operator
panel allowed the system to be enabled and disabled manually.

\begin{figure}[tpb]
\begin{minipage}[b]{0.47\linewidth}
\centering
\includegraphics[width=\columnwidth]{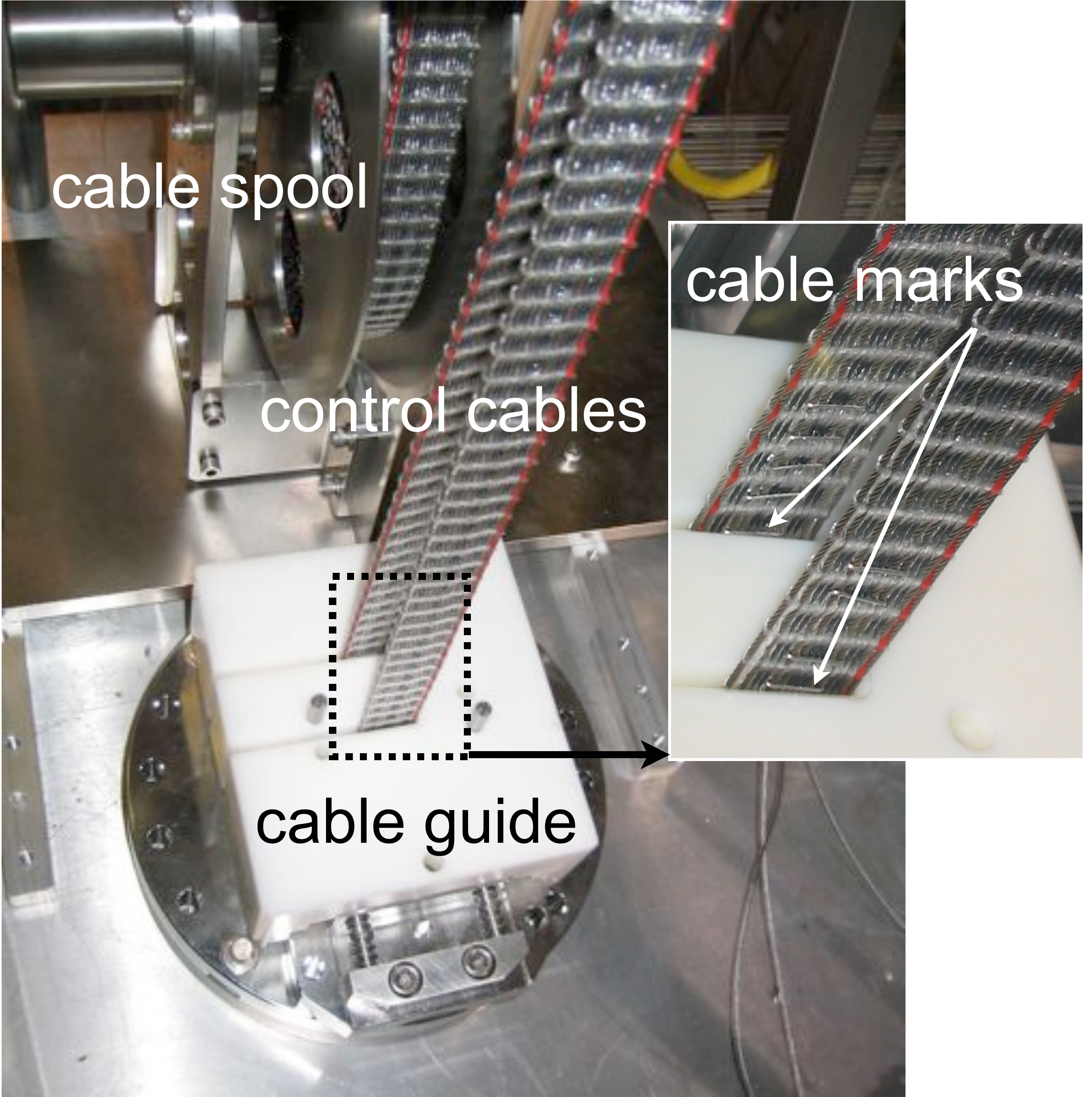}
\caption{The control cables as they passed through the cable guide during operation: A slotted Teflon block was used as a cable guide to minimize vibrations in the cables and cover the access flange.   Also visible are the stainless cable marks which indicated the cable length in binary code and were used as a visual indication of the system's position.}
\label{fig:cable_guide}
\end{minipage}
\hspace{0.5cm}
\begin{minipage}[b]{0.47\linewidth}
\centering
\includegraphics[width=\columnwidth]{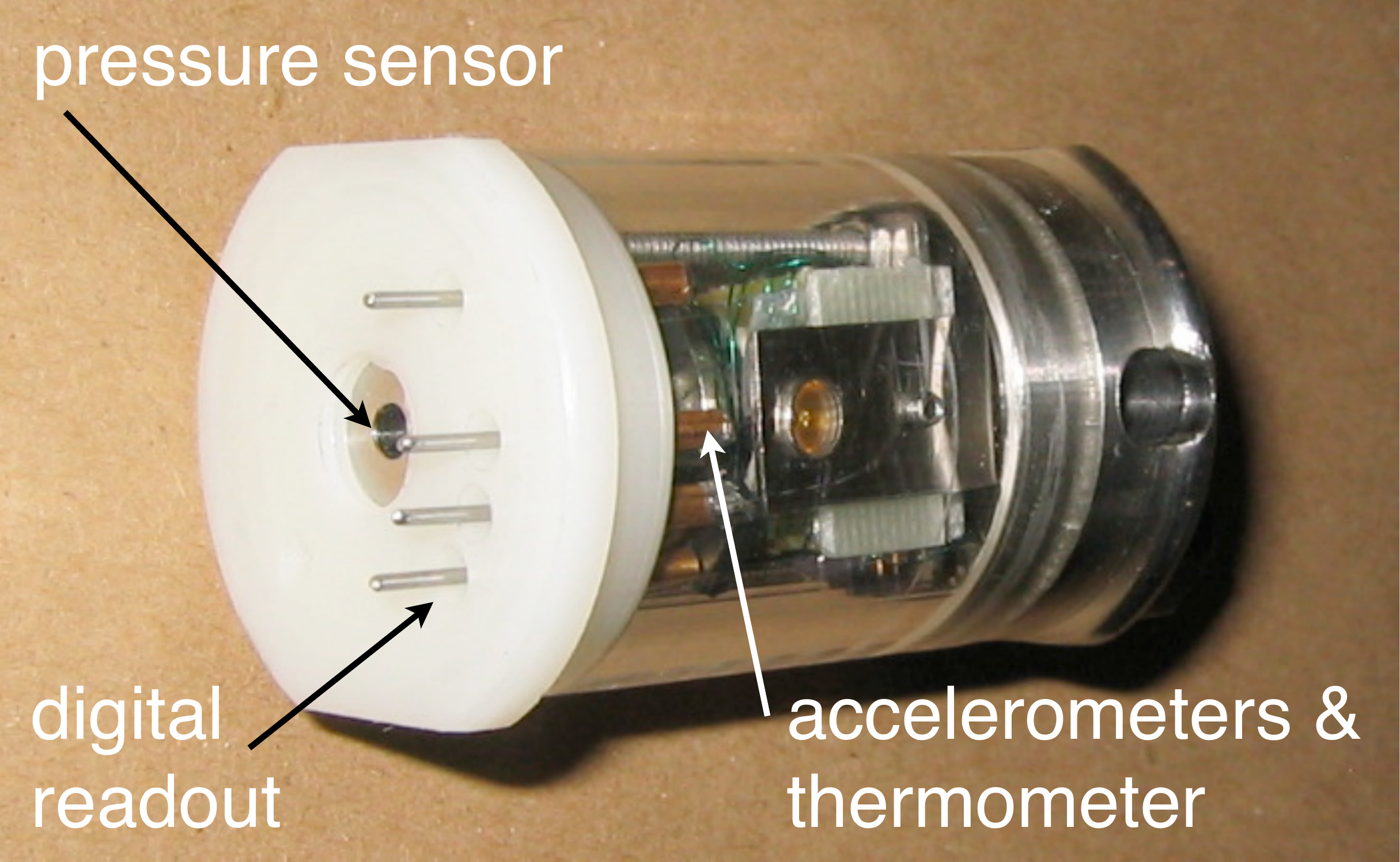} 
\vskip 1.25cm
\caption{A prototype ~\iu\ with acrylic body: The acrylic body was replaced with a stainless-steel body for a more rugged design and to improve the materials compatibility with LS. A pressure sensor, 3-axis accelerometer, and a thermometer are integrated into the unit and read out through a custom-designed connector and 1-Wire communication protocol.}
\label{fig:IUDiagonal}
\end{minipage}
\end{figure}

Direct visual monitoring of the cable spooling was performed by the operators during any movement of the off-axis system. The cable marks, which indicated the absolute length of spooled cable, were used by operators to continually cross check the values in the control program during cable movements.  A cable guide, which covered the open chimney portal in the glovebox, served as a precise reference point for this comparison; it is shown in Fig.~\ref{fig:cable_guide}. 

\begin{figure*}[th]
\begin{center}
\begin{tabular}{ccc}
\includegraphics[scale=0.23, angle=90]{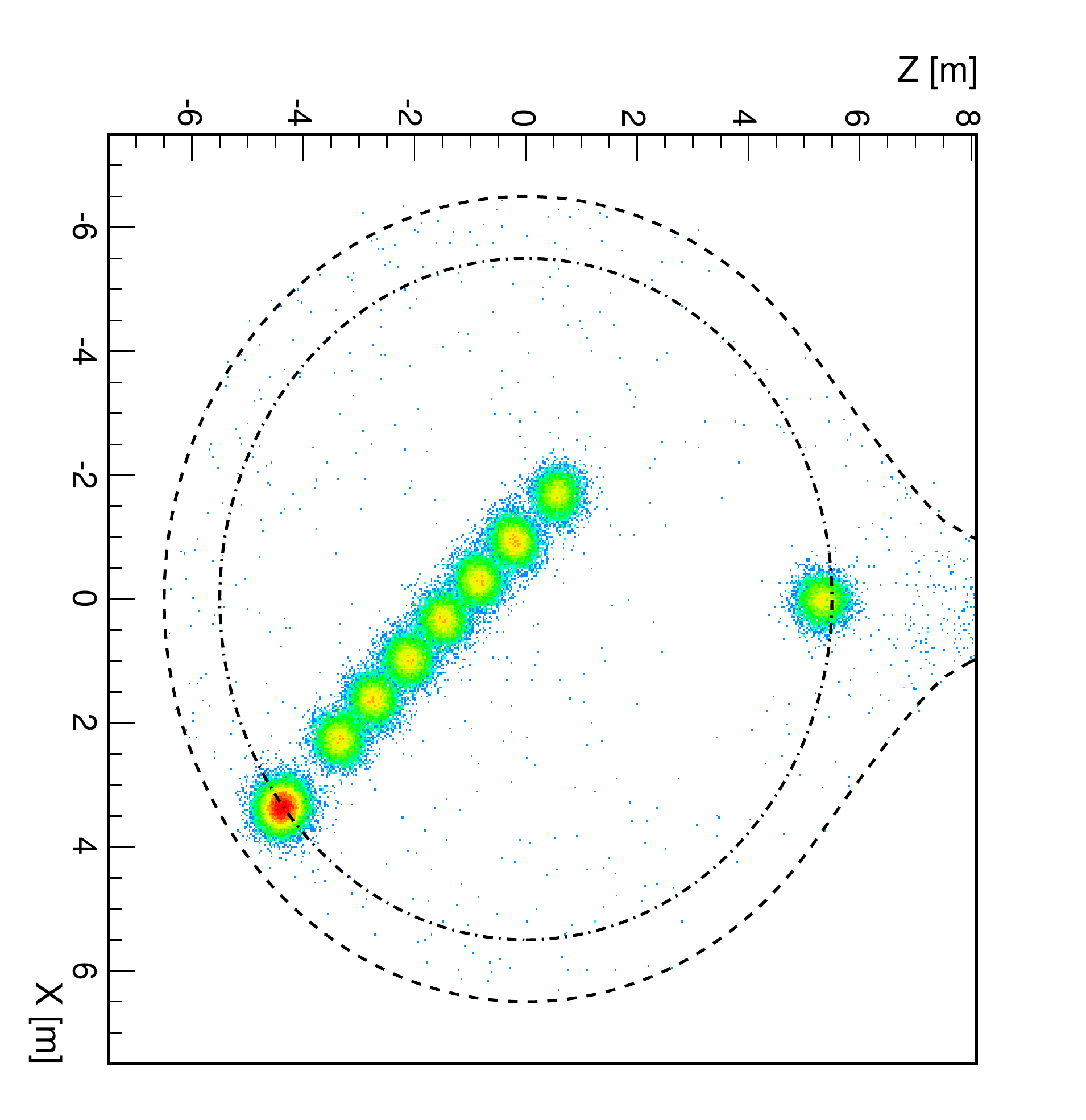} &
\includegraphics[scale=0.23, angle=90]{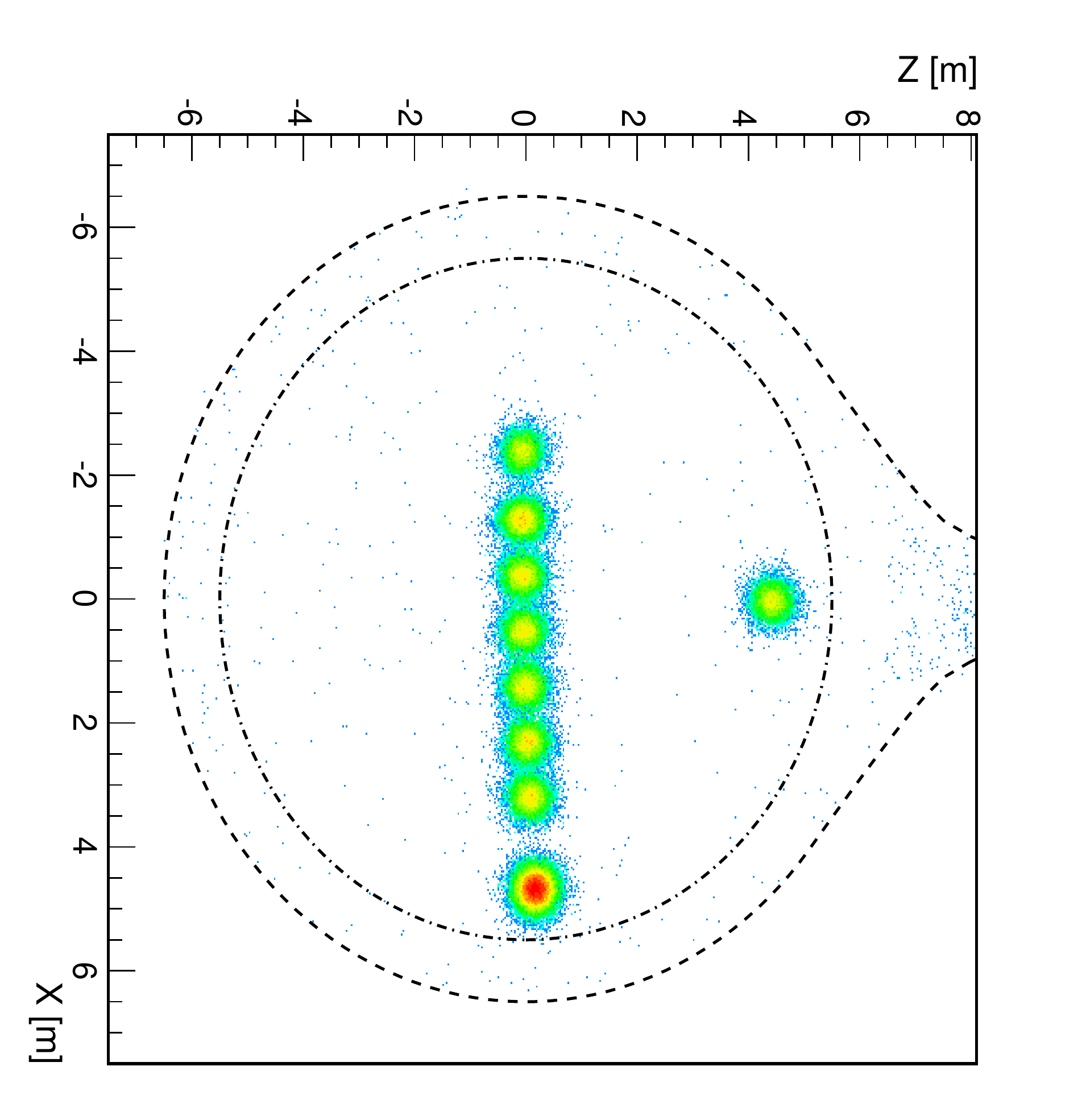} & 
\includegraphics[scale=0.23, angle=90]{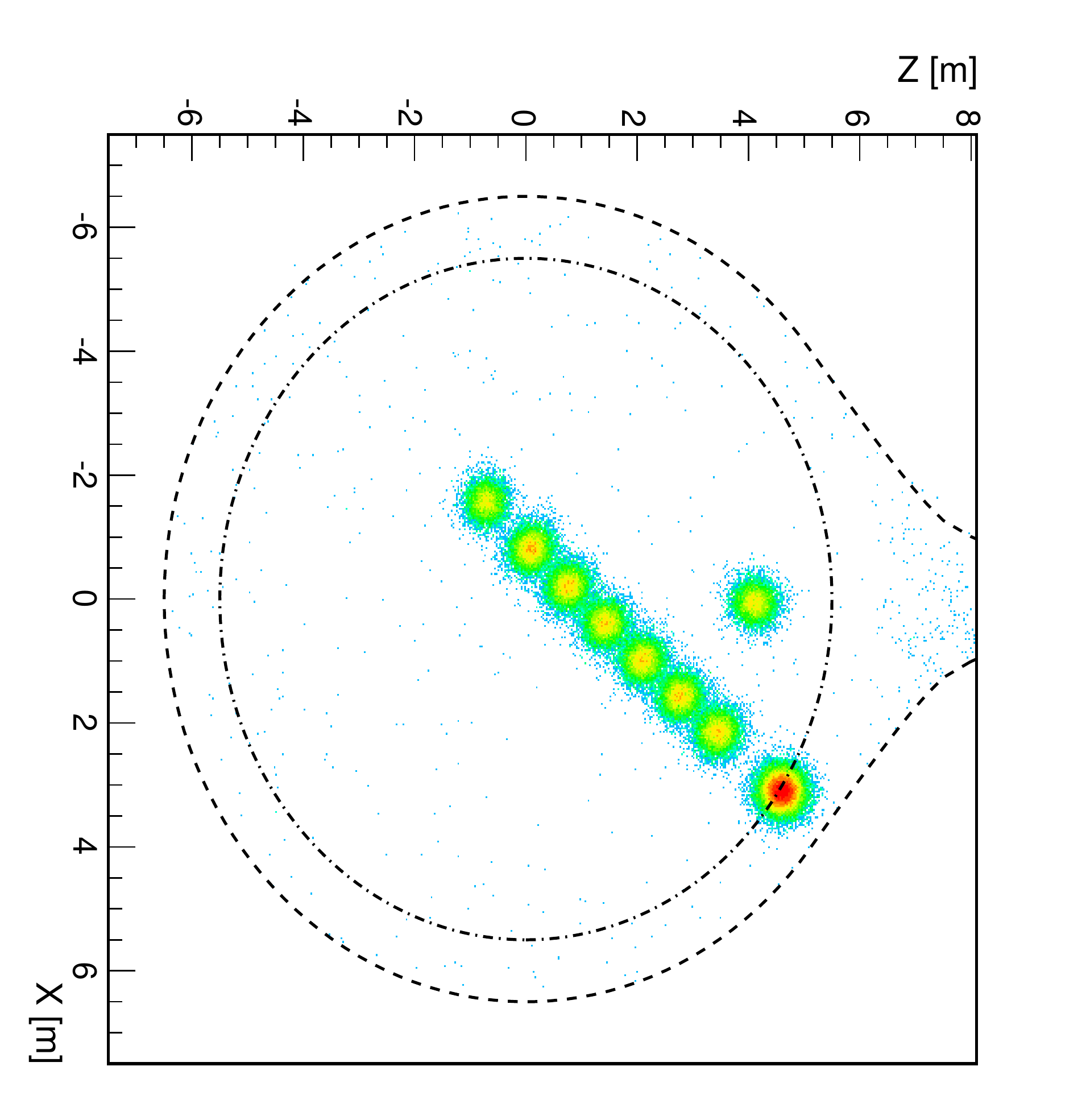} 
\end{tabular}
\caption{The reconstructed position of the radioactive sources in an azimuthal plane of the detector: The colors correspond to the level of detected activity.  The source positions traced the outline of the calibration pole and indicated the location of the pivot block, thus providing confirmation of the system's location.  The outer dotted line represented the balloon boundary.  The inner dot-dashed line was the safety zone which defined the minimum distance between the pole and the balloon. Similar plots were used during the deployment to confirm the location of the system before moving to the next position.  The progression from left to right illustrates the sequence in which the pole was swept through a single azimuthal slice of the detector. }
\label{fig:online_recon}
\end{center}
\end{figure*}

Even with the correct cable length values, one additional failure point for the control program would be the accidental input of the wrong pole geometry. Fast turn-around in the data processing provided an absolute check of the calibration position.  A typical data-taking period at a particular pole position lasted for 15-45 minutes.  During this time, reconstruction algorithms were run on the data as it was collected.  The activity of the sources traced the outline of the calibration pole and indicated the position of the pivot block, as seen in Fig.~\ref{fig:online_recon}.  Events were selected for the energy of the source.  The two-dimensional position in the plane of the pole was plotted to provide confirmation that the system was where it was expected to be. The offline reconstruction did not prevent the actual movement of the pole outside of the safety region.  Hence, the offline reconstruction was always used to verify the pole position when it was still in a well-determined safe region before moving to a more-extended reach zone. 


An ~\iu\ was developed for real-time monitoring of the calibration pole's position independently of the cable lengths and pole geometry.  The ~\iu\ was equipped with a pressure transducer, IC Sensors model 85, for depth measurement and a high-precision digital temperature sensor, Maxim-Dallas DS18S20, to monitor the temperature of the unit. Two infrared LEDs were integrated into the unit to allow for the tracking of the pole with existing CCD cameras in the detector. A one-axis and a two-axis accelerometer, Analog Devices ADXL103CE and  ADXL203CE respectively, were included in the design to monitor the angle of the pole.  The electronics all used the Maxim-Dallas 1-Wire communication protocol.  The 1-Wire protocol gives each device a unique identification number to communicate over one wire, in addition to a common ground. The thermometer was a 1-Wire device, as was the DS2450 four-input analog-to-digital converter, which was used to read out the accelerometers and the pressure transducer. 

The original design had an acrylic body between a metal and nylon endcap, as shown in Fig.~\ref{fig:IUDiagonal}.  This was replaced with a stainless-steel body when the acrylic housing was found to develop incipient cracks. The nylon endcap contained the electrical connections. The stainless-steel endcap provided the channel for the locking pin and a Swagelok fitting for connection to a helium leak detector.

In the end, the \ius\ were not used as a real-time position cross-check because electrical pickup and noise made the 1-Wire communication protocol unreliable.  Despite efforts to better isolate the electronics, crosstalk from the motor power persisted. The pressure sensor data were analyzed offline and found to agree with other measurements within 5~cm provided that the ~\iu\ had sufficient time to come into thermal equilibrium with the surrounding LS. In order to achieve better performance, more care would need to be taken with the temperature compensation of the pressure sensor and the heat conductivity between the thermometer and the pressure sensor. The accelerometer data were also analyzed offline and found to agree with the expected angles within 10$^{\circ}$. In order to achieve better performance with the accelerometers, more care is needed with the mounting and calibration of these devices.

\begin{figure}[tpb]
\begin{minipage}[b]{0.47\linewidth}
\centering
\includegraphics[angle=90, width=1.15\columnwidth]{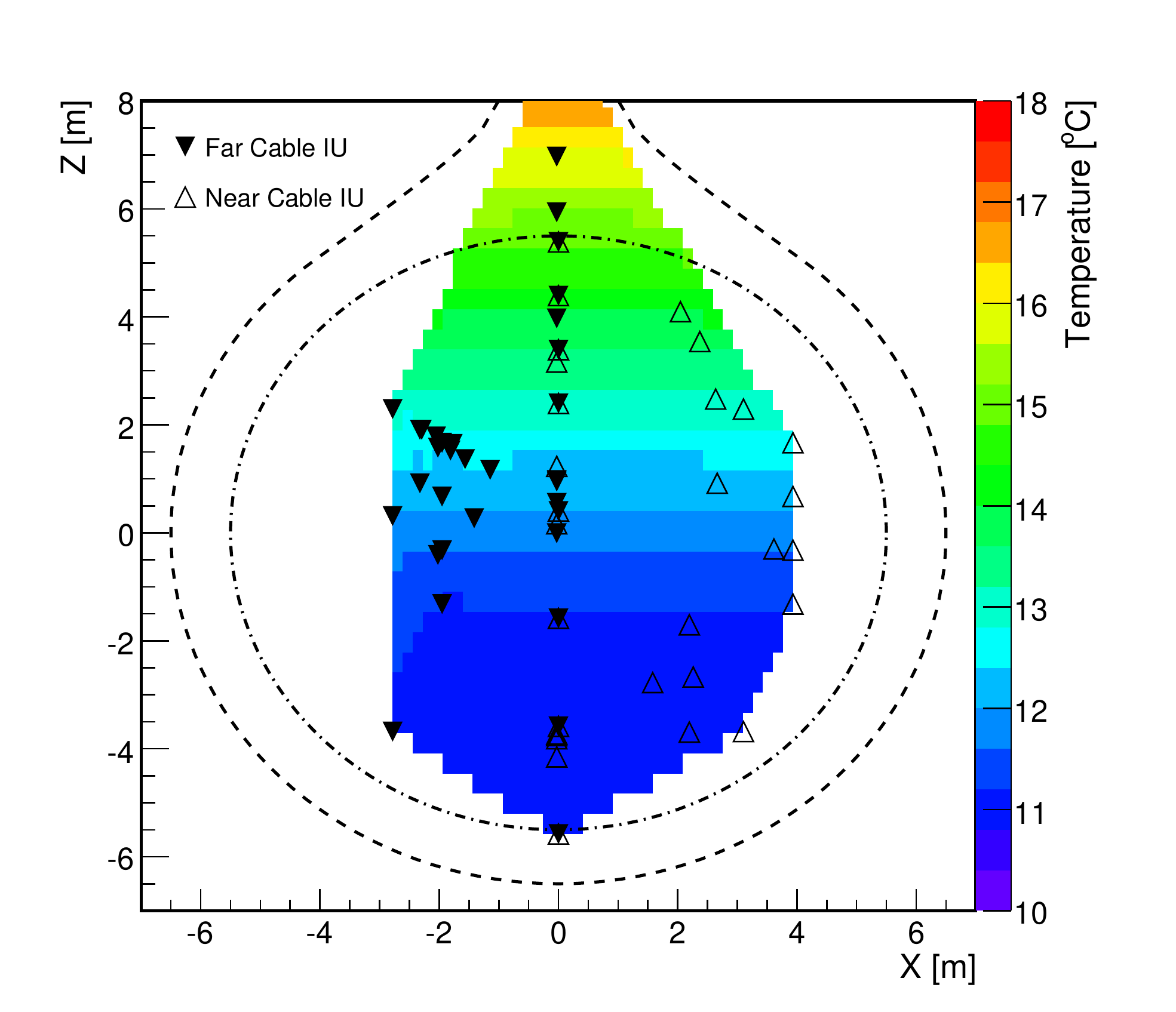}
\caption{Temperature distribution inside KamLAND:  Measurements with the \ius\ at the near and far cable attachments during the detector calibration provided an important map of the temperature layering within the LS.}
\label{fig:IUKamTemp}
\end{minipage}
\hspace{0.5cm}
\begin{minipage}[b]{0.47\linewidth}
\centering
\includegraphics[scale=0.38]{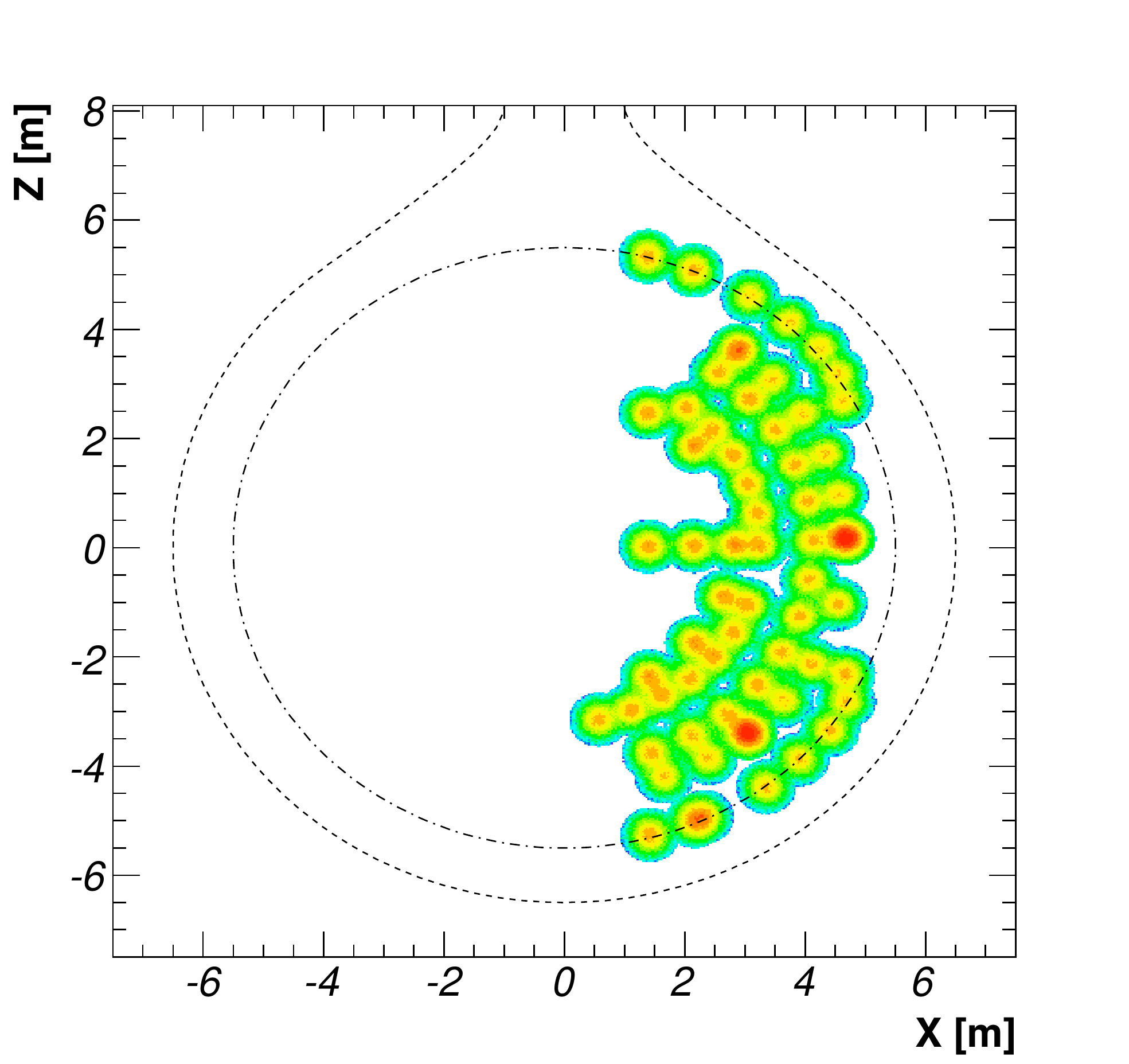}
   \caption{Radial distribution of all off-axis source deployments with the composite $^{60}$Co$^{68}$Ge source excluding initial commissioning runs: This figure illustrates the full radial reach of the calibration system in a single azimuthal plane.}
    \label{fig:sourcedeploy}
    \end{minipage}
\end{figure}

Data from the \ius\ taken during the off-axis deployments provide a detailed measurement of the temperature profile in the detector's LS volume, shown in Fig.~\ref{fig:IUKamTemp}. These data have become very important recently for understanding the temperature layering of the detector. The stability of this temperature gradient is critical to the success of the LS purification for KamLAND's high-purity phase.

\subsection{Offline Position Determination}
\label{sec:pd}


In a separate calculation from the one performed by the motion control system, the geometry of the pole and its suspension system were used to determine the source position to an accuracy of several centimeters.  The center of gravity for an idealized N+1 segment pole suspended from two weightless cables was first calculated.  The cable lengths and the distance between the attachment points was used to calculate the shape and orientation of the pole-cable triangle.  The source-end position was then specified by the pole angle and the distance along the pole as measured from the center of gravity, which lay along the vertical line passing through the suspension point.  
  
\begin{figure}[tpb]
\begin{center}
\includegraphics[width = 9cm, clip, trim=10mm 30mm 10mm 30mm]{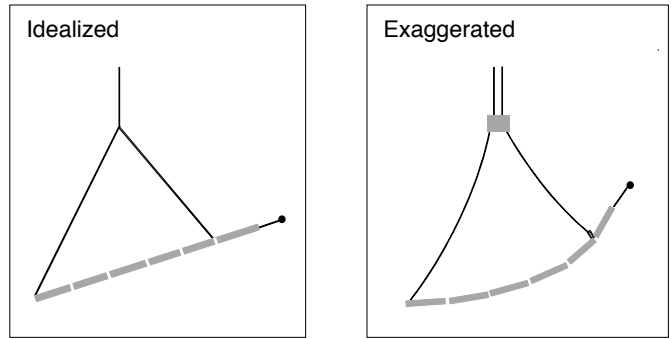}
\caption{The left schematic shows an idealized calibration pole geometry.  The right panel illustrates corrections to the ideal case, such as pole and cable deflections, kinks in the pole joints, and the finite size of the pivot block.  The effects are exaggerated for clarity. }
\label{fig:ideal_real_geom}
\end{center}
\end{figure}

Next, various correction factors were applied to account for center-of-gravity shifts caused by the weight of the cable, pole and cable deflections, buoyancy, and geometric modifications attributable to the pivot block and the attachment points.  See Fig.~\ref{fig:ideal_real_geom} for the ideal and exaggerated-real geometry of the calibration device.

We found that the weight of the cables produced horizontal shifts of $-2$ to $+12$~cm and the deflection of the pole under its own weight and the tensions in the cables introduced shifts in the pole positions of 0 to 8~cm.  The effective horizontal position of the suspension point could shift by $\pm1.5$~cm depending on the tensions in the cables, which were constrained to a separation of 3~cm above the pivot block.  The exact value of the shifts depended on the total length of the pole and the angle of the pole relative to the horizontal.  Geometric corrections, primarily related to the pivot block, produced additional 1 to 2~cm shifts.  Sagging in the control cables introduced only small vertical shifts that were typically a few millimeters.   Buoyancy effects in the LS produced approximately 1~cm changes in the pole deflection and the position of the center of gravity.  A welding error in the mounting of the source-end pole segment and the source mount resulted in visible kinks, which displaced the source-end by 1-2~cm. 

\begin{figure}[htpb]
\begin{center}
\includegraphics[angle=90, width=0.7\columnwidth]{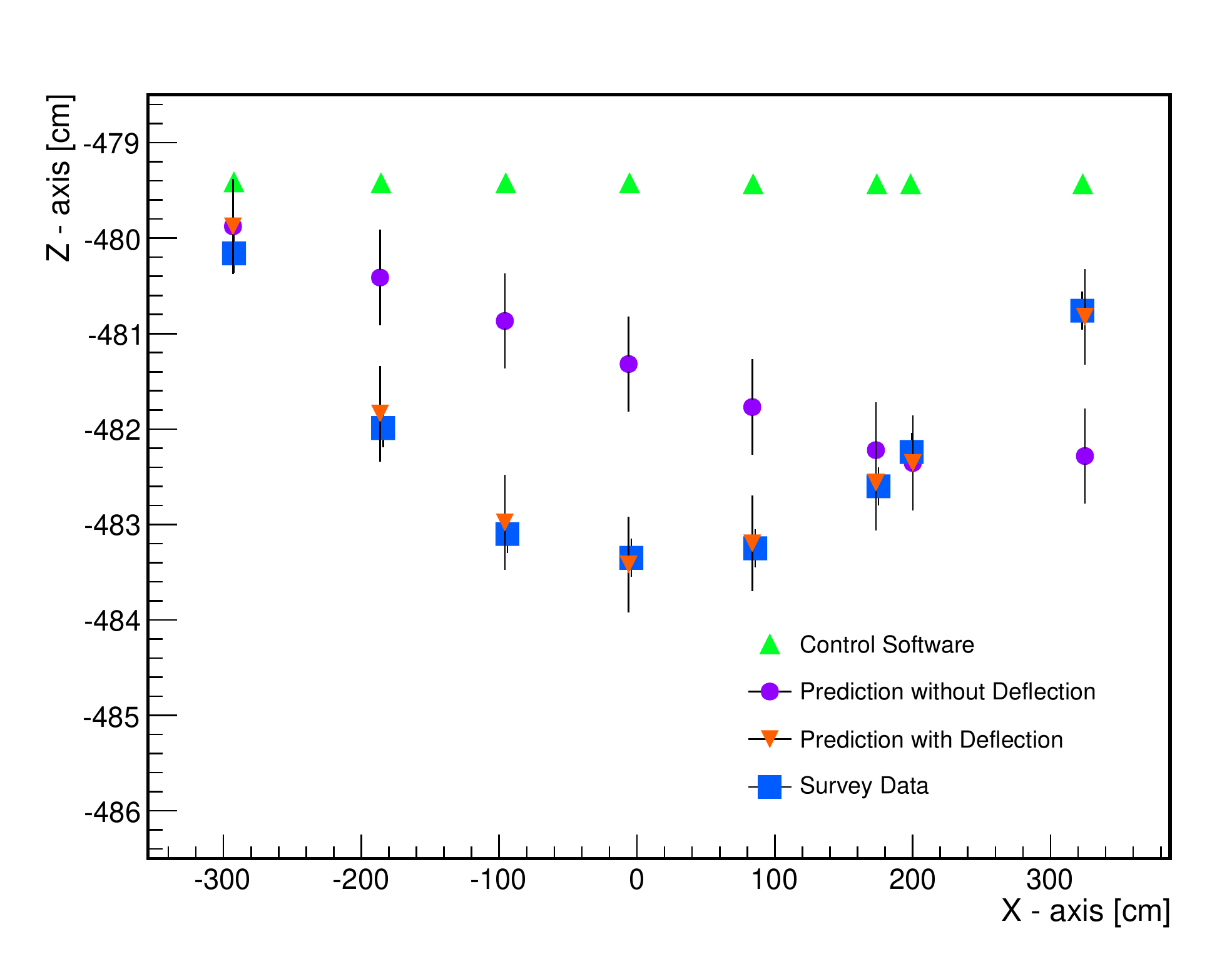}
\caption{Comparison of survey positions determined during the commissioning of the system with calculated calibration pole positions. }
\label{fig:survey_compI}
\end{center}
\end{figure}

After these corrections were applied, the overall positioning accuracy of the source-end was determined to $\pm2$~cm.  These calculations were compared with detailed survey studies, which were performed on the suspended pole system in the high bay of the LBNL Bevatron.  See Fig.~\ref{fig:survey_compI} for a comparison in the case of a horizontal \fiveplusone\ pole.  The results agree well when the pole deflection correction is included.  




\subsection{Data Collection}
\label{sec:data_collection}

In total, more than 500 hours of detector live time were utilized to perform the off-axis calibrations. Excluding the initial commissioning period, data were taken at over 60 unique positions, in a single azimuthal plane with a composite \composite\ source.  The azimuthal plane mapping was performed with the \fourplusone, \fiveplusone, \fiveplusonew, and \sixplusonew~configurations.  For each pole configuration, data were taken at a range of zenith positions while pivoting the pole about the detector center.  This motion mapped  out a series of semicircles\footnote{This was done in order to make the best use of the relative distance between the embedded sources and the source tip in the study of position deviations. The most efficient motion for moving the device was actually a direct translation in the up or down direction, which maps out a cylindrical surface.  }.  The configurations and their corresponding radial reach are summarized in Table~\ref{tab:deploy_table}.  Additional points were probed along the detector equatorial plane and near the polar regions by translating the calibration pole up and down. At the equator of the detector volume, the radial reach of the longest pole configuration was 4.6~m.  Except for a 2~m band around the equator of the detector, the vertical translations of the pole allowed the maximum radial reach of 5.5~m to be accessible.  

\begin{table}[!t]
\begin{center}
\vskip 0.1in
\begin{tabular*}{\columnwidth}{@{\extracolsep{\fill}}  l  l  }\hline
\hline
Calibration Pole Configuration & Radial Reach at Equator \\\hline
     3+1      &  2.5~m \\
     4+1      &  2.8~m \\
     5+1      &  3.3~m \\
     5+1W     &  4.1~m \\
     6+1W     &  4.6~m \\\hline
\hline
\end{tabular*}
\end{center}
\caption{Reach of all deployed calibration pole configurations: The reach is defined as the radial position of the source-end when the pole is in the equatorial plane of the detector. }
\label{tab:deploy_table}
\end{table}

The total set of all calibration positions is shown in Fig.~\ref{fig:sourcedeploy}.  Using the \composite\ source, a sparser set of points were taken at three additional azimuthal positions 90$^\circ$ apart. The reconstructed data show little variation in azimuthal position, so this was deemed adequate. To study the energy dependence of reconstruction deviations, deployments were also made to a subset of the points using other sources.  

\begin{table}[!t]
\begin{center}
\vskip 0.1in
\begin{tabular*}{\columnwidth}{@{\extracolsep{\fill}} l l r}
\hline
\hline
Source                 & Activity~[Bq]     &    Calibration Energy \\\hline                  
$^{68}$Ge$^{60}$Co     & 200/120           &    $\gamma$ (1.022, 2.506~MeV) \\
$^{203}$Hg	       & 1300              &    $\gamma$ (0.279~MeV) \\
$^{241}$Am$^9$Be       & 1.85$\times10^{6}$            &    $\gamma$ (4.4~MeV), n ($<$10~MeV)   \\
$^{210}$Po$^{13}$C     & 3$\times 10^6$ ($^{210}$Po)   &    $\gamma$ (6.13~MeV), n ($<$7.5~MeV) \cite{PoC} \\\hline
\hline
\end{tabular*}
\end{center}
\caption{Radioactive sources used in calibration: This table shows the activity visible to the detector and does not necessarily reflect the complete activity of the listed isotope.}
\label{tab:source_table}
\end{table}

\section{Performance \& Post-Deployment Analysis}
\label{sec:performance}

\subsection{Position Determination, Stability and Reproducibility}


An initial portion of the data-taking phase was devoted to commissioning each of the pole configurations.  During these runs, the pole was fully instrumented with the pin sources and the composite \composite\ source.   We used the reconstructed calibration data to study the position of the pole during the test operations.   The primary use of these data was to determine the predictability and reproducibility of the pole positions.  This testing period was also important for establishing safe operation procedures and to characterize the motion of the system while inside the detector.  

Despite the implementation of flat cables, the test deployments revealed that the calibration pole twisted about its center-of-mass axis as the control cables were spooled out.   This effect was most notable and persistent for the lightest pole configuration, \threeplusone.  However, all configurations larger than \threeplusone\ stabilized quickly in $\phi$ (azimuthal position) once the cables stopped moving. Though the $\phi$ resting-point of a pole configuration could not be predicted prior to its first deployment, each configuration was found to return to a reproducible $\phi$ position within 10 degrees.   The preferred $\phi$ resting-point was found to be unique to the pole configuration.  The $\phi$ resting point did remain largely fixed as the pole was swept through the zenith angle.  As the detector response was found to be quite symmetric in $\phi$, the inability to control the absolute azimuthal position was not a great concern.  

In the zenith and radial directions, the pole was well-constrained.  The test deployment data show that the 2-D position of the source within the plane of the pole is predictable (to a few cm) and reproducible (to a few mm).  In these spatial directions, the pole was very stable, with no movement observed during data-taking.  

Onsite test deployments revealed a 1~cm systematic uncertainty in the lengths of the cables.  Part of this uncertainty originated from a ``hysteresis'' effect where the retracted length of the cables was found to be greater than the spooled length by 1-2~cm over a 10~m deployment of cable.  Thermal expansion and stress-strain tests on a sample of cable were unable to explain the effect by an order of magnitude.  To minimize this bias, the cables were recalibrated using the known values of the cable marks after every few meters of movement.   An additional source of uncertainty in the cable-lengths arose from the measured distance between cable marks, which was known to better than a millimeter. Because the cumulative length of the cables was determined by summing the distances between cable marks, even a sub-millimeter systematic bias in the distance between cable marks could build up to 1~cm over the total deployed cable length, which was typically 12-18~m.  

With exact knowledge of the system geometry, center-of-mass and cable lengths, we expect a 2-D position uncertainty of 2~cm; as explained in section~\ref{sec:pd}.  However, the additional approximately 1~cm uncertainties in the cable-lengths adds a non-negligible uncertainty to the position of the source.  A study was done to calculate the source-end displacement for 10000 random 1~cm perturbations of one or both cables.   The geometry of the system introduced a lever-arm effect.  This resulted in typical source-end displacements of 2.5-6~cm, mostly in the vertical direction.  The exact magnitude of the displacement depended on the length and orientation of the pole.  Based on these studies, we determined that the mean uncertainty in the absolute position of the source-end was less than 5~cm.



\subsection{Source Shadowing}
\label{sec:data_shadowing}

The off-axis system was considerably larger in both surface area and mass compared to the more extensively used on-axis system.   It was necessary to study the possibility of ``shadowing'' in the off-axis calibration data, which arises from the system blocking and absorbing some scintillation light along with the particles emitted from the source.  To investigate this effect, data were taken along the central axis of the detector by allowing the assembled calibration pole, with \composite\ source, to hang vertically and translating it up and down.  The vertical-pole data were directly compared with extensive \composite\ source data from the simpler, pole-free on-axis system.  This study revealed no systematic bias in energy reconstruction for events at the source-end of the pole at the 0.25\% level.  However, the energy of events from the $^{60}$Co sources embedded along the pole exhibited degradation in measured energy due to the scattering of the emitted gammas inside the extra material of the pole and pin-source encapsulation.  We observed ~2\% shifts in the energy of events from the embedded sources.  The exact size of the shift depended on the orientation of the pole, which suggests that the effect was at least partly due to optical shadowing.  

The vertical pole study indicated no systematic bias in measured position due to shadowing effects.   This is not surprising because the position reconstruction algorithm depends on the time-of-flight of the scintillation photons and is thus largely insensitive to the number of scintillation photons detected (for energy deposits $>1$~MeV).  Simulations~\cite{Geant4} of the calibration system within the detector qualitatively confirmed both the position and energy shadowing results, though the size of the energy degradation for the embedded sources was predicted to be smaller by a factor of two.

\subsection{Data Analysis}
\label{sec:data_analysis}
For the experiment, we are primarily interested in measuring the energy deviation, $\Delta E(r,\theta,\phi)$, and the radial position deviation, $\Delta R(r, \theta, \phi)$ of reconstructed events.  The measured energy deviation is simply the difference between the known energy of the source and the energy measured by the detector.  Each measured energy deviation is associated with the reconstructed position of the source.  We used all the sources listed in Table~\ref{tab:source_table} to measure the energy deviation.  We did not use the pin sources owing to the shadowing effect described above.  

The radial position deviations were measured by comparing the known distance between a pin source and the source-end to the reconstructed distance between these two points.  To measure the radial deviations, we used only the data where a far-end pin source was close to the detector center.   This was necessary because we approximated the radial deviation by equating it to the pin-to-source distance deviation.   This is a good approximation when the pole runs through the center of the detector and the radial deviation is small at the position of the reference pin.  This was the case for pin positions within a few meters of the detector center. 

\begin{figure}[htpb]
\begin{center}
\includegraphics[angle=90, width=0.48\columnwidth]{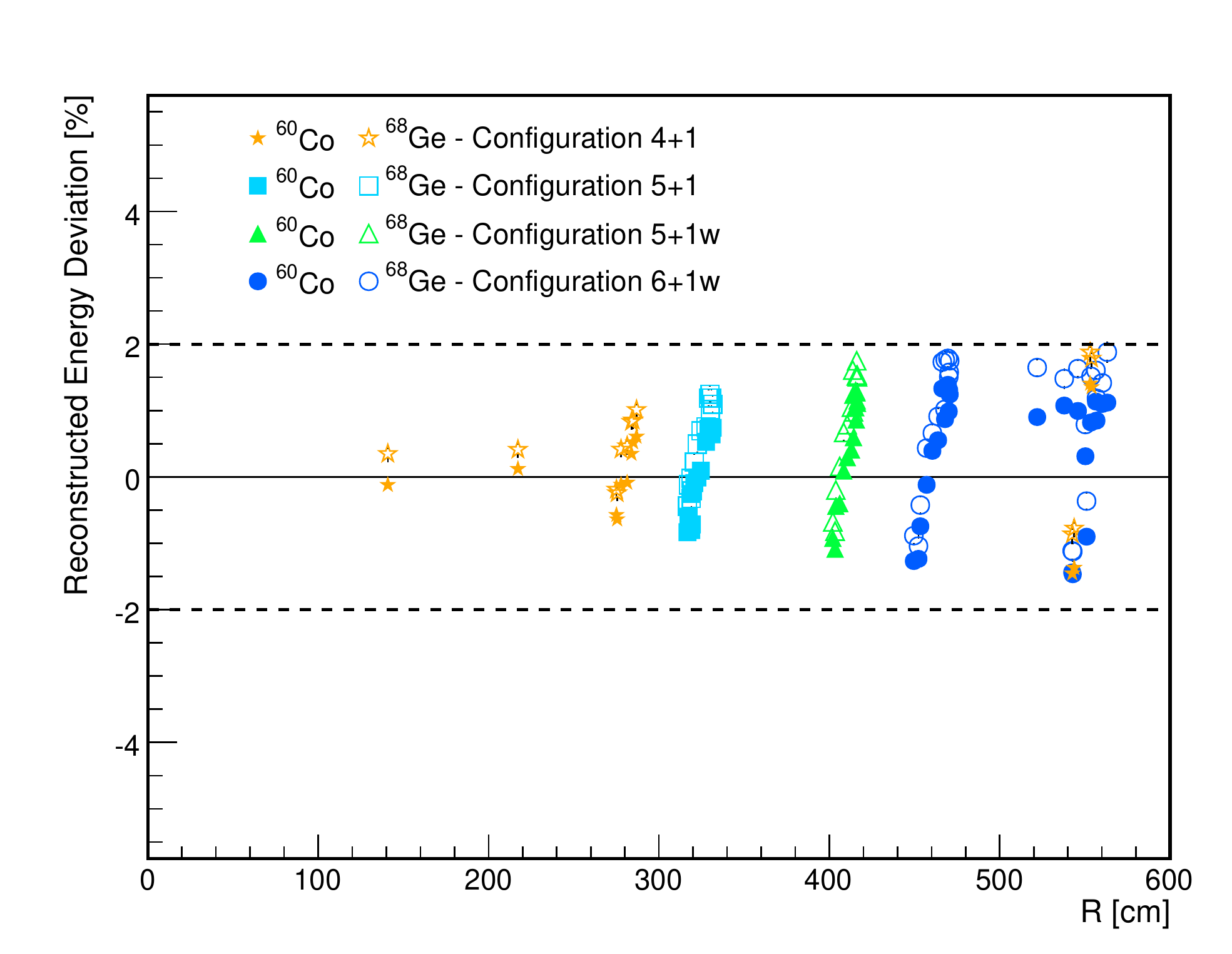}
\includegraphics[angle=90, width=0.48\columnwidth]{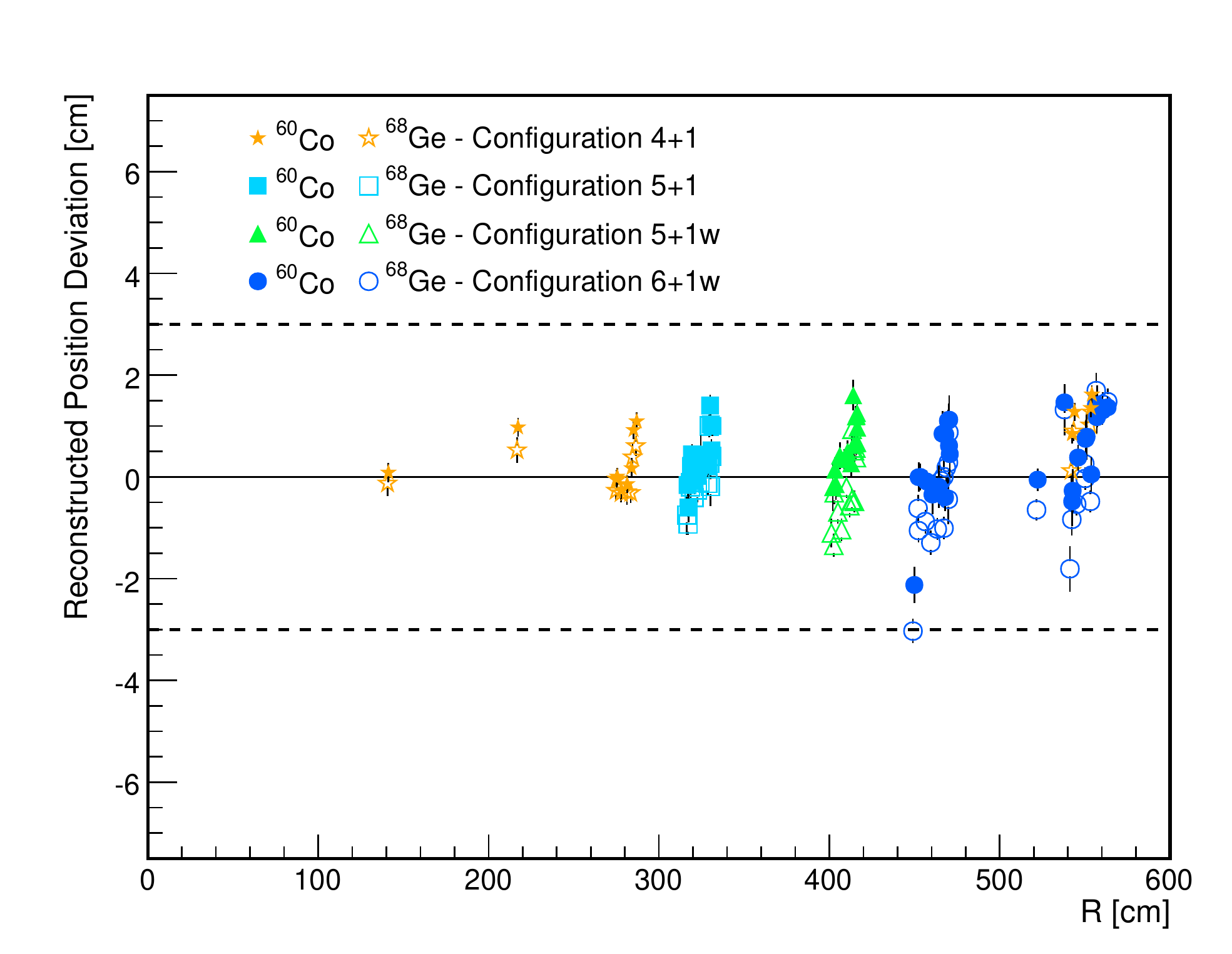}
\caption{The measured reconstruction deviations as a function of detector radius:   The energy deviations were found to be less than 2\% (left). The radial position deviations are less than 3~cm (right).  The different points correspond to a given pole configuration as follows: yellow stars (4+1),  cyan squares (5+1), green triangles (5+1W), blue squares (6+1W).  The data with filled points were measured at the 2.506~MeV peak of $^{60}$Co. The data with hollow points were measured at the 1.022~MeV annihilation gamma peak of $^{68}$Ge.  }
\label{fig:measbias}
\end{center}
\end{figure}

See Fig.~\ref{fig:measbias} for a subset of the energy and radial position deviations measured with the off-axis calibration system.  The measured deviations were found to vary with radius and zenith angle.  The magnitudes of the observed systematic deviations are small, $< 2$\% energy deviation and $< 3$~cm radial position deviation, and they show no significant variation with energy.  The variation of these deviations in azimuthal angle is smaller than the variation in radius and zenith angle, as expected from the detector geometry. The off-axis deviations are within the range of earlier estimates, which were deduced from on-axis data and cosmogenic-induced events.  The azimuthal symmetry of these deviations suggest that they arise from geometrical effects because the detector has imperfect spherical symmetry (imperfections include the tear-drop shape of balloon, support ropes, and non-uniformity of photo-cathode coverage, especially at the top and bottom of the detector).  

The fiducial volume uncertainty was determined using the measured 3~cm upper limit to the radial deviations.   The uncertainty in a spherical volume, defined by the maximum reach of 5.5~m, was estimated as a uniform 3~cm radial deviation at 5.5~m.  This value yielded a fractional uncertainty in the volume of 1.6\%.    A cross-check of this measurement, using spallation-induced events, yielded consistent values but with a larger uncertainty of 4\%~\cite{KLprl3}.\footnote{The spallation events were used to estimate the fiducial volume prior to the commissioning of this calibration system\cite{KLprl1,KLprl2}.  With this method, the volume was calculated by computing the ratio of cosmogenic-induced $^{12}$B/$^{12}$N $\beta$-decays reconstructed inside of the fiducial volume to all reconstructed $^{12}$B/$^{12}$N.  This ratio was then multiplied by the total volume of LS as measured with flow-meters during the detector filling.  Because the volume ratio is measured at the $^{12}$B/$^{12}$N energies, which is higher than the \antinu\ spectrum, the precision of this method is limited by the uncertainty in the energy dependence of the volume ratio as measured with these decays (2.7\%).  The precision is further limited by the uncertainty in the total volume measured by the flowmeters (2.1\%). }  

In the most recent \antinu\ analysis, the fiducial volume was extended 0.5~m past the maximum reach of the system.  The remaining volume between the radii of 5.5~m and 6~m was calculated using the spallation event method$^3$.  When the 5.5~m volume, measured by the calibration system, was combined with the extension to 6~m, measured with the spallation events, a fiducial volume uncertainty of 1.8\% was assigned to the total volume \cite{KLprl3}. This represents a reduction of more than a factor of 2 over previous levels of uncertainty that were obtained with the spallation analysis alone.


\section{Conclusion}
\label{sec:conclusion}

\subsection{Design Evaluation}
\label{sec:lessons}

As this was the first system of its kind to be built and operated, we summarize here the unexpected challenges we overcame and point out potential areas of improvement for the design of future systems.  The \ius\ exhibited several unexpected problems, which rendered them ineffective as real-time monitors of the system position (see section~\ref{sec:pm}). In particular, the sensitivity to electrical noise from the spool motors caused the 1-Wire protocol to fail whenever the pole was in motion.   The cable-mark watching protocol was developed to compensate, using human effort, for what was lost in the instrumentation.  In general, this was an adequate solution, however it significantly complicated movement of the calibration pole because it required visual inspection of the cables whenever the system was in motion.  Thus, the lack of real-time instrumentation on the pole had a profound effect on both the complexity of procedures necessary to operate the system and the number of person-hours that were required.   



We experienced several problems from the woven design of the control cables.  The unexplained systematic effects which caused the cable lengths measured by the encoders to differ by several cm from the true values slightly compromised our ability to determine the absolute position of the calibration source (see section~\ref{sec:performance}).   Originally, a 2~cm accuracy in absolute position was required in order to achieve our goal in fiducial volume measurements, but the actual result that we obtained was more than twice this.   In the end, the implementation of the pin sources was extremely successful because it enabled a circumvention of this problem.   We also note that the flat cables, while they represented an improvement over cables with a circular cross-section, did not completely eliminate the problem that the system position was poorly constrained in the azimuthal direction. Fortunately this was not a matter of concern for KamLAND, but in future designs where the azimuthal position may matter, this must be taken into consideration.  Finally, we note that the cables were susceptible to \radon\ plate-out.  We believe it is a result of the material choice and the extra surface area of the weave.  Though the cable met our cleanliness requirements for the reactor \antinu\ phase, the level to which it attracts \radon\ daughters makes it unsuitable for the high-purity phase of KamLAND.

\subsection{Summary}
\label{sec:summary}

We have successfully designed and operated a calibration device for the KamLAND detector.  This device allowed for the positioning of radioactive sources throughout a target volume within 5.5~m of the detector center.  Though simple in concept, stringent requirements on radioactive cleanliness,  compatibility with the LS, and safety of the detector necessitated a careful and intricate design.   Precision motion control and strict operator procedures enabled this device to be safely assembled and manipulated within the fragile inner volume of KamLAND.  The intrinsic radioactive contamination from the system was found to be small but could become problematic once KamLAND achieves significantly increased purity levels.  Using data taken with the calibration system, the fiducial volume was determined to 1.6\% uncertainty at a radius of 5.5~m. This data has reduced the fiducial volume uncertainty -- the dominant systematic uncertainty on the rate of \antinu s detected --  by a factor of two.  Ongoing analysis of the calibration data will also help improve the reconstruction of data in the high purity phase of KamLAND.


\vskip 0.2in
\noindent
{\bf Acknowledgements}
\vskip 0.1in
\noindent

We would like to thank several colleagues for their support and help in the design and construction of this system: Al Smith was always helpful with our material counting needs and shared much advice in the handling of low-background components.  We thank Fred Beiser for advice on a variety of electronic issues and John Wolf for helping us construct and debug the \ius.  Leo Greiner provided valuable insight and information based on his experiences with the previous KamLAND on-axis calibration system.  We are grateful to George Webber for his help in the initial prototyping stages of this project and in the fabrication of the pivot block and cable clamps.  We thank Mark Rosen for his work on the conceptual design of the system.

The KamLAND experiment is supported by the Japanese Ministry of Education, Culture, Sports, Science and Technology, and under the United States Department of Energy Office grant DEFG03-00ER41138 and other DOE grants to individual institutions.  The Kamioka Mining and Smelting Company has provided services for activities in the mine. 






\end{document}